\documentclass[onecollarge]{svjour2}       
\smartqed  
\usepackage{graphicx}
\usepackage{amssymb}
\usepackage{amsmath}
\usepackage{mathtools}
\usepackage{diagrams}
\usepackage{multirow}
\usepackage{pbox}
\usepackage{enumerate}

\usepackage{caption}
\DeclareCaptionLabelFormat{adja-page}{#1 #2 \emph{(previous page)}}
\usepackage{subfig}
\makeatletter
\DeclareRobustCommand*\subref{\@ifstar\sf@@subref\sf@subref}
\makeatother

\usepackage{bm}
\newcommand{\eq}[1]{Eq.\ (\ref{#1})}
\newcommand{\fig}[2]{Fig.\ \ref{#1}#2}

\newcommand{\var}{{\rm var}}
\newcommand{\cov}{{\rm cov}}
\newcommand{\bea}{\begin{eqnarray}}
\newcommand{\eea}{\end{eqnarray}}
\newcommand{\beas}{\begin{eqnarray*}}
\newcommand{\eeas}{\end{eqnarray*}}
\newcommand{\data}{{\rm data}}
\newcommand{\real}{{\rm real}}

\newcommand{\const}{{\rm const}}

\newcommand{\set}[1]{\left\{ #1 \right\}}

\newcommand{\braket}[1]{\left\langle #1 \right\rangle}

\begin{document}
\title{Learning quantitative sequence-function relationships from massively parallel experiments
}

\titlerunning{Mutual information vs.\ likelihood}        

\author{ Gurinder S. Atwal \and Justin B. Kinney}
\institute{J. B. Kinney \at
              Simons Center for Quantitative Biology, Cold Spring Harbor Laboratory \\
              Cold Spring Harbor, NY 11724\\
              Tel.: +1-516-3675230\\
              \email{jkinney@cshl.edu}           
           \and
           G. S. Atwal \at
              Simons Center for Quantitative Biology, Cold Spring Harbor Laboratory 
}

\date{Received: date / Accepted: date}
\maketitle

\begin{abstract}

A fundamental aspect of biological information processing is the ubiquity of sequence-function relationships -- functions that map the sequence of DNA, RNA, or protein to a biochemically relevant activity. Most sequence-function relationships in biology are quantitative, but only recently have experimental techniques for effectively measuring these relationships been developed. The advent of such ``massively parallel'' experiments presents an exciting opportunity for the concepts and methods of statistical physics to inform the study of biological systems. After reviewing these recent experimental advances, we focus on the problem of how to infer parametric models of sequence-function relationships from the data produced by these experiments. Specifically, we retrace and extend recent theoretical work showing that inference based on mutual information, not the standard likelihood-based approach, is often necessary for accurately learning the parameters of these models. Closely connected with this result is the emergence of ``diffeomorphic modes'' -- directions in parameter space that are far less constrained by data than likelihood-based inference would suggest. Analogous to Goldstone modes in physics, diffeomorphic modes arise from an arbitrarily broken symmetry of the inference problem. An analytically tractable model of a massively parallel experiment is then described, providing an explicit demonstration of these fundamental aspects of statistical inference. This paper concludes with an outlook on the theoretical and computational challenges currently facing studies of quantitative sequence-function relationships. 

\end{abstract}

\section{Introduction}

\begin{figure}[t] 
   \centering
   \includegraphics{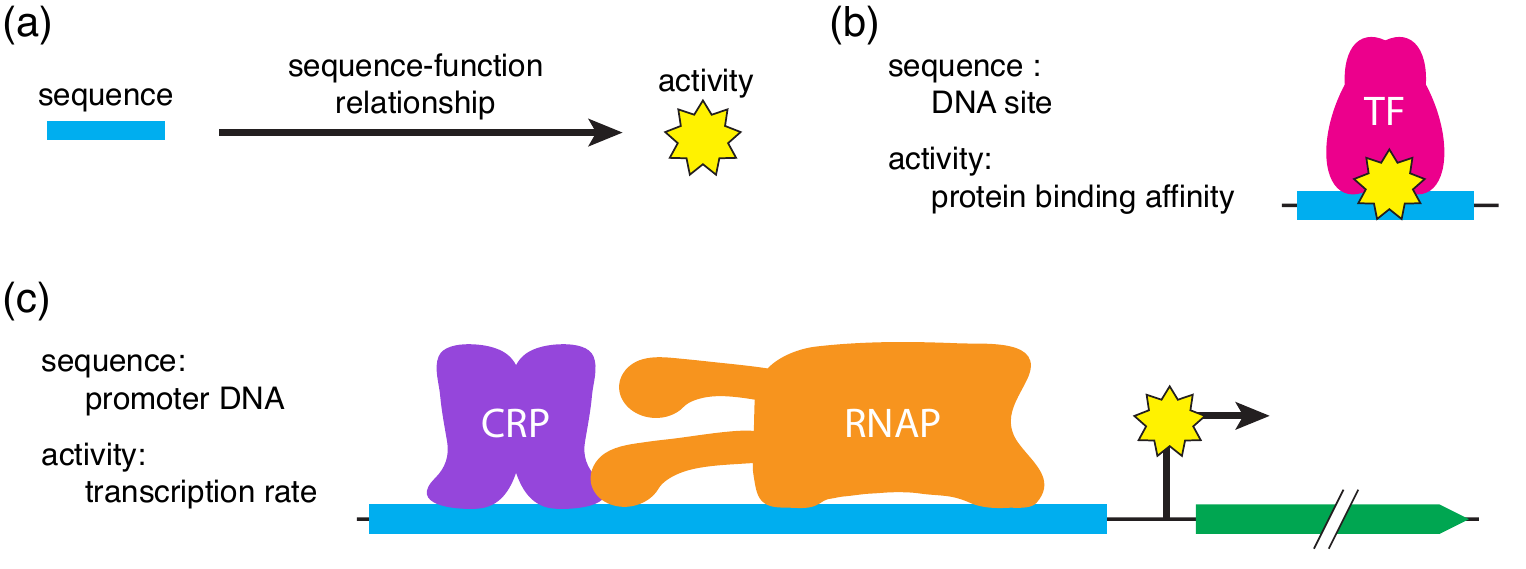} 
   \caption{Sequence-function relationships in biology. (a) A sequence-function relationship maps a biological sequence (blue bar) to a biologically relevant activity (yellow star). (b) One of the simplest sequence-function relationships is how the affinity (star) of a transcription factor protein (magenta) for its DNA binding site depends on the sequence of that site (blue). (c) A more complicated sequence-function relationship describes how the rate of mRNA transcription depends on the DNA sequence of a gene's promoter region. At the \emph{lac} promoter of \emph{E. coli} (illustrated), this transcription rate  (star) depends on how strongly both the transcription factor CRP (purple) and the RNA polymerase holoenzyme (RNAP; orange) bind their respective sites within the promoter region (blue).}
   \label{fig:relationships}
\end{figure}

A major long-term goal in biology is to understand how biological function is encoded within the sequences of DNA, RNA, and protein. The canonical success story in this effort is the genetic code: given an arbitrary sequence of messenger RNA, the genetic code allows us to predict with near certainty what peptide sequence will result. There are many other biological codes we would like to learn as well. How does the DNA sequence of a promoter or enhancer encode transcriptional regulatory programs? How does the sequence of pre-mRNA govern which exons are kept and which are removed from the final spliced mRNA? How does the peptide sequence of an antibody govern how strongly it binds to target antigens? 

A major difference between the genetic code and these other codes is that while the former is qualitative in nature, the latter are governed by sequence-function relationships that are inherently quantitative. Quantitative sequence-function relationships\footnote{These have also called quantitative sequence-activity maps, or QSAMs \cite{Melnikov:2012dw}.} describe any function that maps the sequence of a biological heteropolymer to a biologically relevant activity (Fig.\ 1a). Perhaps the simplest example of such a relationship is how the affinity of a transcription factor protein for its DNA binding site depends on the DNA sequence of that site (Fig.\ 1b). Such relationships are a key component of the  more complicated relationship between the DNA sequence of a promoter or enhancer (which typically binds multiple proteins) and the resulting rate of mRNA transcription (Fig.\ 1c). In both of these cases, the activities of interest (affinity or transcription rate) can vary over orders of magnitude and yet still be finely tuned by adjusting the corresponding sequence (binding site or promoter/enhancer). Similarly other sequence-function relationships, like the inclusion of exons during mRNA splicing or the affinity of a protein for its ligand, are fundamentally quantitative. 

The study of quantitative sequence-function relationships presents an exciting opportunity for the concepts and methods of statistical physics to shed light on biological systems. There is a natural analogy between biological sequences and the microstates of physical systems, as well as between biological activities and physical Hamiltonians. Yet we currently lack answers to basic questions a statistical physicist might ask, such as ``what is the density of states?'' or ``is a relationship convex or glassy?" The answers to such questions may well have important consequences for diverse fields including biochemistry, systems biology, immunology, and evolution. 

Experimental methods for measuring sequence-function relationships have improved dramatically in recent years. In the mid 2000s, multiple ``high-throughput'' methods for measuring the DNA sequence specificity of transcription factors were developed; these methods include protein binding microarrays (PBMs) \cite{Mukherjee:2004um,Berger:2006wp}, \emph{E.\ coli} one-hybrid technology (E1H) \cite{Meng:2005dp}, and microfluidic platforms \cite{Maerkl:2007tv}. The subsequent development and dissemination of ultra-high-throughput DNA sequencing technologies then led, starting in 2009, to the creation of a number of ``massively parallel'' experimental techniques for probing a wide range of sequence-function relationships (Table 1). These massively parallel assays can readily measure the functional activity of $10^3$ to $10^8$ sequences in a single experiment by coupling standard bench-top techniques to ultra-high-throughput DNA sequencing.

\renewcommand{\arraystretch}{1.3}
\begin{table}
\begin{tabular}{|c|c|c|c|l|}
\hline
\textbf{sequence}				& \textbf{activity}		& \textbf{system}		&	\textbf{name}		& \textbf{publication} \\ \hline \hline
\multirow{5}{*}{\parbox{2cm}{\centering DNA \\ binding \\ sites}}  
						& \multirow{5}{*}{\parbox{2cm}{\centering protein-DNA \\ binding \\ affinity}}		
												& \multirow{5}{*}{\parbox{2cm}{\centering purified protein}}			
																		& Bind-n-Seq	& Zykovich et al., 2009 	\cite{Zykovich:2009dx} \\ \cline{4-5}
						& 						& 						& HT-SELEX		& Zhao et al., 2009 			\cite{Zhao:2009gs} \\ \cline{4-5}
						& 						& 						&				& Jolma et al., 2010 		\cite{Jolma:2010hz} \\ \cline{4-5}
						& 						& 						& EMSA-Seq		& Wong et al., 2011 			\cite{Wong:2011gi} \\ \cline{4-5}
						& 						& 						& SELEX-Seq	& Slattery et al., 2011 	\cite{Slattery:2011hxa} \\ \hline 
\multirow{6}{*}{\parbox{2cm}{ \centering promoter/\\enhancer \\ DNA}}			
						& \multirow{6}{*}{\parbox{2cm}{\centering transcription \\ rate}}	
												& purified protein	&				& Patwardhan et al., 2009 	\cite{Patwardhan:2009cw} \\ \cline{3-5}
						& 						& bacteria				& Sort-Seq		& Kinney et al., 2010 		\cite{Kinney:2010tn} \\ \cline{3-5}
						& 						& cell culture		& MPRA			& Melnikov et al., 2012 	\cite{Melnikov:2012dw} \\ \cline{3-5}
						& 						& mouse liver			& 				& Patwardhan et al., 2012 	\cite{Patwardhan:2012hy} \\ \cline{3-5}
						&						& yeast				&				& Sharon et al., 2012		\cite{Sharon:2012io} \\ \cline{3-5}
						& 						& mouse retina		& CRE-Seq		& Kwasniesk et al., 2012 	\cite{Kwasnieski:2012hu} \\ \hline 
						
\multirow{5}{*}{\parbox{2cm}{ \centering protein}}
						& ligand binding			& phage display	& DMS			& Fowler et al., 2010 		\cite{Fowler:2010gt} \\ \cline{2-5}
						& cellular growth rate	& yeast			& EMPIRIC		& Hietpas et al., 2011 		\cite{Hietpas:2011bp} \\ \cline{2-5}
						& toxin activity			& bacteria			&				& Adkar et al., 2012			\cite{Adkar:2012ea}	\\ \cline{2-5}
						& H1N1 binding			& yeast display	& 				& Whitehead et al., 2012	\cite{Whitehead:2012jt} \\ \cline{2-5}
						& GPCR expression			& bacteria			& 				& Schlinkmann et al., 2012	\cite{Schlinkmann:2012il} \\ \hline 
\multirow{4}{*}{\parbox{2cm}{ \centering RNA}}
						& mRNA translation		& bacteria			& 				& Holmqvist et al., 2013 	\cite{Holmqvist:2013hx} \\ \cline{2-5}
						& sRNA targeting			& bacteria			& qSortSeq		& Peterman et al., 2014		\cite{Peterman:2014cd}  \\ \cline{2-5}
						& mRNA translation		& cell culture	& 				& Oikonomou et al., 2014 	\cite{Oikonomou:2014goa} \\ \cline{2-5}
						& mRNA translation 		& cell culture 	& FACS-Seq		& Noderer et al., 2014 		\cite{Noderer:2014fj}	\\ \hline
						
replication origins		& DNA replication		& yeast			& ARS-Seq		& Liachko et al., 2013  	\cite{Liachko:2013jc} \\  \hline

endonuclease sites 		& DNA cutting			& purified protein	& 				& Thyme et al., 2014			\cite{Thyme:2014hna} \\ \hline

\end{tabular}
\caption{Massively parallel experiments used for studying various sequence-function relationships. Columns show the type of sequences interrogated, the sequence activity assayed, the biological system on which the experiments were performed, the name (if any) of the experimental technique, and the publication first describing the method. This table is not exhaustive; it only describes some of the earliest experiments in each type of system. }
\end{table}

Massively parallel experiments are very unlike conventional experiments in physics: they are typically very noisy and rarely provide direct readouts of the quantities that one cares about. Moreover, the noise characteristics of these measurements are difficult to accurately model. Indeed, such noise generally exhibits substantial day-to-day variability. Although standard inference methods require an explicit model of experimental noise, it is still possible to precisely learn quantitative sequence-function relationships from massively parallel data even when noise characteristics are unknown \cite{Kinney:2007dh,Kinney:2014ge}. 

The ability to fit parametric models to these data reflects subtle but important distinctions between two objective functions used for statistical inference: (i) likelihood, which requires {\it a priori} knowledge of the experimental noise function and (ii) mutual information \cite{Cover:1991ti}, a quantity based on the concept of entropy, which does not require a noise function. In contrast to the conventional wisdom that more experimental measurements will improve the model inference task, the standard maximum likelihood approach will \emph{typically} never learn the right model, even in the infinite data limit, if one uses an imperfect model of experimental noise. Model inference based on mutual information does not suffer from this ailment. 

Mutual-information-based inference is unable to pin down the values of model parameters along certain directions in parameter space known as ``diffeomorphic modes'' \cite{Kinney:2014ge}. This inability is not a shortcoming of mutual information, but rather reflects a fundamental distinction between how diffeomorphic and nondiffeomorphic directions in parameter space are constrained by data. Analogous to the emergence of Goldstone modes in particle physics due to a specific yet arbitrary choice of phase, diffeomorphic modes arise from a somewhat arbitrary choice one must make when defining the sequence-dependent activity that one wishes to model. Likelihood, in contrast to mutual information, is oblivious to the distinction between diffeomorphic and nondiffeomorphic modes.

We begin this paper by briefly reviewing a variety of massively parallel assays for probing quantitative sequence-function relationships. We then turn to the problem of learning parametric models of these relationships from the data that these experiments generate. After reviewing recent work on this problem \cite{Kinney:2014ge}, we extend this work in three ways. First, we show that ``diffeomorphic modes'' of the parametric activity model that one wishes to learn are ``dual'' to certain transformations of the corresponding model of experimental noise (the ``noise function''). This duality reveals a symmetry of the inference problem, thereby establishing a close analogy with Goldstone modes. Next we compute and compare the Hessians of likelihood and mutual information. This comparisons suggests an additional analogy between this inference problem and concepts in fluid mechanics. Finally, we work through an analytically tractable model of a massively parallel experiment of protein-DNA binding. This example explicitly illustrates the differences between likelihood- and mutual-information-based approaches to inference, as well as the emergence of diffeomorphic modes.

It should be noted that the inference of receptive fields in sensory neuroscience is another area of biology in which mutual information has proved useful as an objective function, and that work in this area has also provided important insights into basic aspects of machine learning \cite{Paninski:2003eq,Sharpee:2004ty,Sharpee:2006vb,Kouh:2009ik,Rajan:2011vi}. Indeed, the problem of learning quantitative sequence-function relationships in molecular biology is very similar to the problem of learning receptive fields in neuroscience \cite{Kinney:2014ge}. The discussion of this problem in the neuroscience context, however, has largely avoided in-depth analyses of how mutual information relates to likelihood, as well as of how diffeomorphic modes emerge. 

\section{Massively parallel experiments probing sequence-function relationships}

\begin{figure}[p] 
  \centering
  \includegraphics{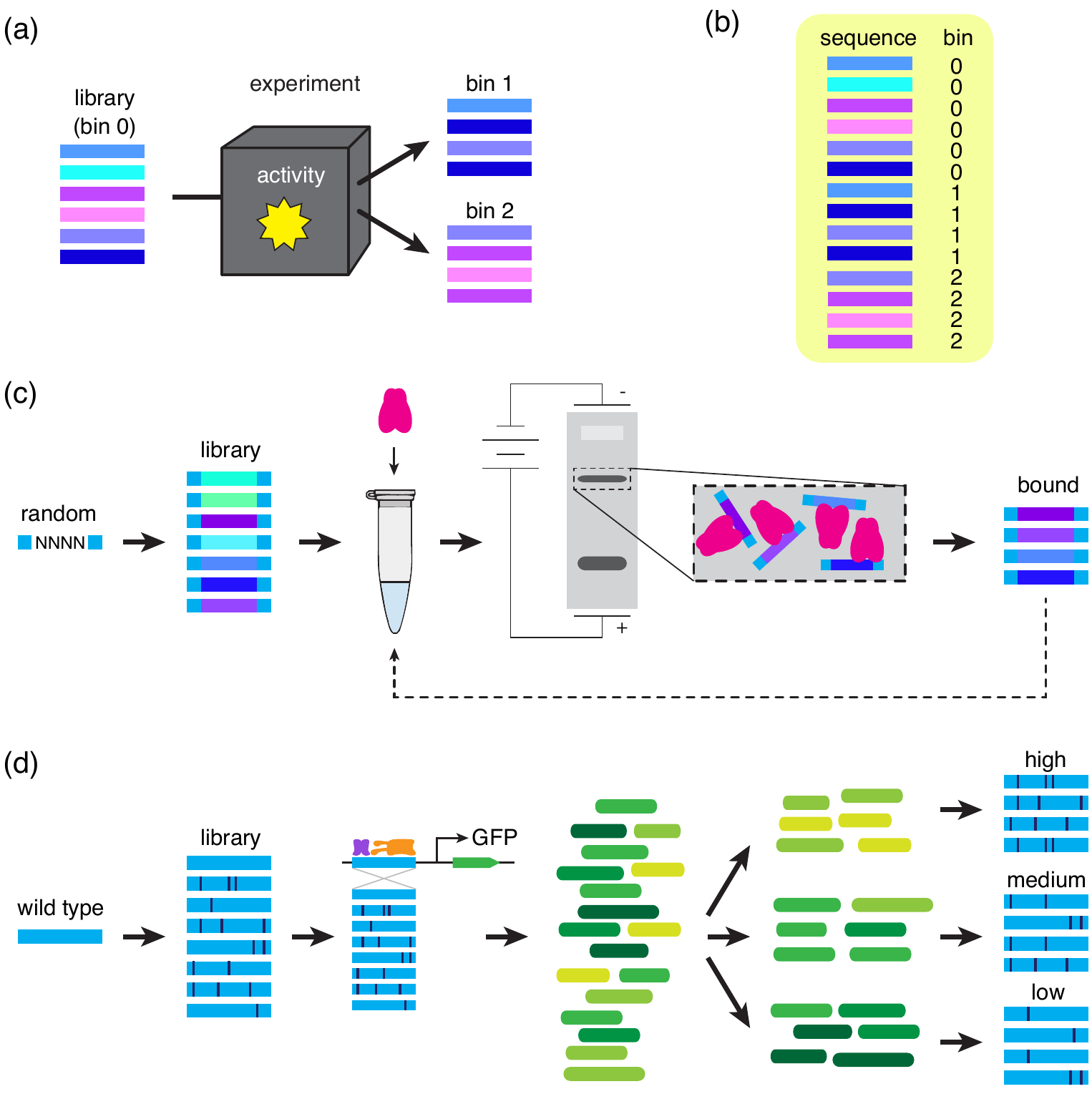}\\
  \caption{Overview of massively parallel experiments for studying quantitative sequence-function relationships. (a) The input to each experiment is a library of different sequences that one wishes to test. The output is one or more bins of sequences; each sequence in each bin is randomly selected from the library with a weight that depends on a measurement of that sequence's activity (star). (b) The resulting data set consists of a list of (non-unique) sequences, each sequence assigned to either the input library or one of the output bins. (c) Illustration of experimental methods for measuring the sequence-dependent binding energy of purified transcription factor proteins. The input library typically consists of random DNA flanked by constant sequence. This library DNA is mixed with the protein of interest and binding is allowed to come to equilibrium. DNA bound by protein is then separated from unbound DNA, e.g.\ by running complexes on a gel (shown), then sequenced along with a sample from the input library.  (d) Sort-Seq \cite{Kinney:2010tn} is a massively parallel experiment that uses a library of partially mutagenized sequences to probe the mechanisms of transcriptional regulation employed by a specific wild type promoter of interest.  Mutant promoters are cloned upstream of the GFP gene, and \emph{E.\ coli} cells harboring these expression constructs are sorted into bins using FACS. The mutant promoters in each bin, as well as promoters from the input library, are then sequenced.}
  \label{fig:experiments}
\end{figure}

\begin{figure}[t] 
   \centering
   \includegraphics{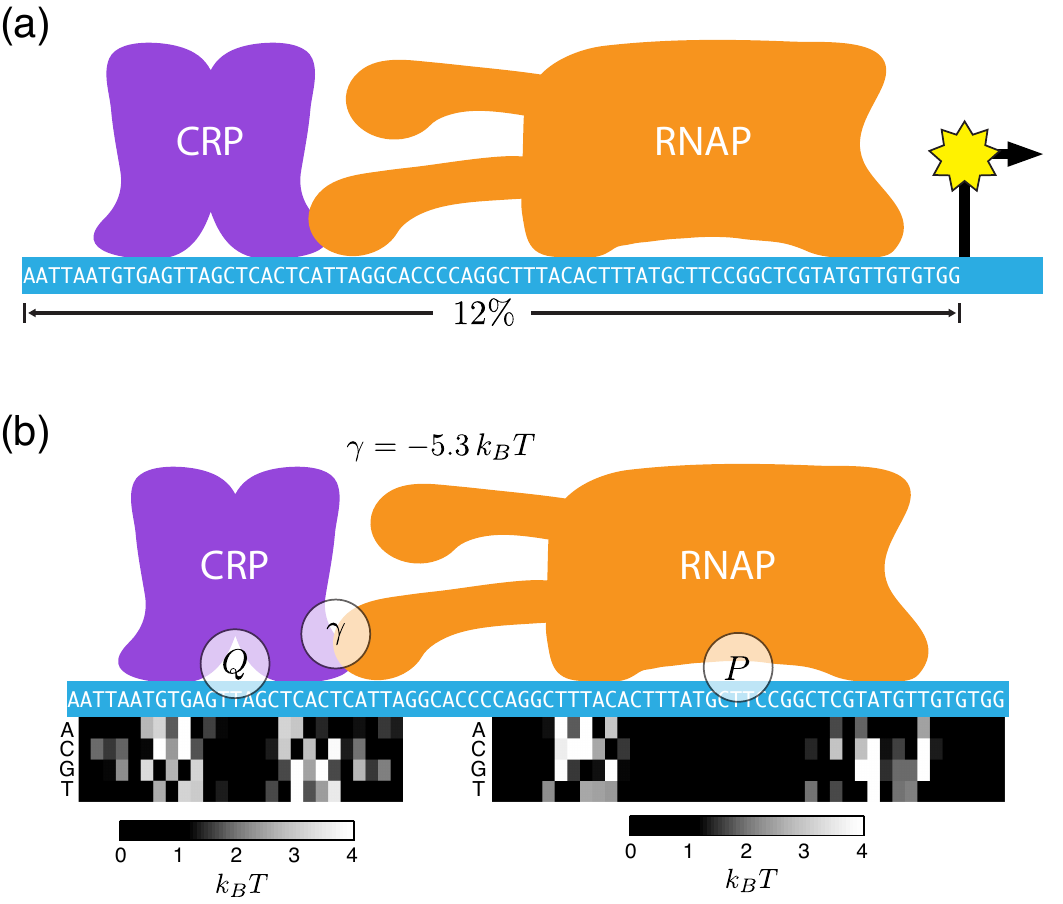} 
   \caption{The \emph{lac} promoter region studied in \cite{Kinney:2010tn}. (a) Sort-Seq was used to dissect a 75 bp region of the \emph{E.\ coli lac} promoter using a library consisting of wild type sequences mutagenized at 12\% per nucleotide, i.e., each library sequence had 9 mutations on average. (b) The resulting data were used to learn a quantitative sequence-function relationship, the mathematical form of which reflected an explicit biophysical model of transcriptional regulation. This model included two ``energy matrices'' describing the sequence-dependent binding energy of CRP ($Q$) and RNAP ($P$) to their respective sites. It also included a value for the interaction energy $\gamma$ between these two proteins.}
 \label{fig:transcription}
\end{figure}

All of the massively parallel experiments in Table 1 share a common structure (Fig.\ 2a). The first step in each experiment is to generate a large set of (roughly $10^3$ to $10^8$) different sequences to measure. This set of sequences is called the ``library.'' Multiple different types of libraries can be used depending on the application. One then performs an experiment that takes this library as input, and as output provides a set of one or more ``bins'' of sequences. Each output bin contains sequences selected from the library with a weight that depends on the measured activity of that sequence. Finally, a sample of sequences from each of the output bins, as well as from the input library, are determined using ultra-high-throughput DNA sequencing. The resulting data thus consists of a long list of (typically non-unique) DNA sequences, each assigned to a corresponding bin (Fig.\ 2b). It is from these data that we wish to learn quantitative models of sequence-function relationships. 

Some of the earliest massively parallel experiments were designed to measure the specificity of  purified transcription factors for their DNA binding sites \cite{Zykovich:2009dx,Zhao:2009gs,Jolma:2010hz,Wong:2011gi,Slattery:2011hxa} (Fig.\ 2c). The library used in such studies consists of a fixed-length region of random DNA flanked by constant sequences used for PCR amplification.  This library is mixed with the transcription factor of interest, after which protein-bound DNA is separated from unbound DNA, e.g., by running the protein-DNA mixture on a gel. Protein-bound DNA is then sequenced, along with the input library. 

Using a library of random DNA to assay protein-DNA binding has the advantage that the same library can be used to study each protein. This is particularly useful when performing assays on many different proteins at once (e.g., \cite{Jolma:2010hz,Jolma:2013fh}). On the other hand, only a very small fraction of library sequences will be specifically bound by the protein of interest. Moreover, because proteins typically bind DNA in a non-specific manner, such experiments are often performed serially  in order to achieve substantial enrichment.\footnote{This serial enrichment approach is known as SELEX and is much older than ultra-high-throughput DNA sequencing; see \cite{Oliphant:1989vb,Tuerk:1990ux,Ellington:1990ct,Blackwell:1990tq,Wright:1991uw}.} 

The first massively parallel experiment to probe how multi-protein-DNA complexes regulate transcription in living cells was Sort-Seq \cite{Kinney:2010tn} (Fig.\ 2d). The sequence library used in this experiment was generated by introducing randomly scattered mutations into a ``wild-type'' sequence of interest, specifically, the 75 bp region of the promoter of the \emph{lac} gene in \emph{E.\ coli} depicted in Fig.\ 3a. A few million of these mutant promoters were cloned upstream of the green fluorescent protein (GFP) gene. Cells carrying these expression constructs were grown under conditions favorable to promoter activity and were then sorted into a small number of bins according to each cell's measured fluorescence. This partitioning of cells was accomplished using fluorescence-activated cell sorting (FACS) \cite{Herzenberg:1976th}, a method that can readily sort  $\sim10^4$ cells per second. The mutant promoters within each sorted bin as well as within the input library were then sequenced, yielding measurements for $\sim 2 \times 10^5$ variant promoter sequences. We note that advances in DNA sequencing have since made it possible to accumulate much more data, and it is no long difficult to measure the activities of $\sim 10^7$ different sequences in this manner. 

Massively parallel experiments using partially mutagenized sequences provide data about sequence-function relationships within a localized region of sequence space centered on the wild type sequence of interest. Measuring these local relationships can provide a wealth of information about the functional mechanisms of the wild type sequence. For instance, the Sort-Seq data of \cite{Kinney:2010tn} allowed the inference of an explicit biophysical model for how CRP and RNAP work together to regulate transcription at the \emph{lac} promoter (Fig.\ 3b). In particular, the authors used their data to learn quantitative models for the \emph{in vivo} sequence specificity of both CRP and RNAP. Model fitting also enabled measurement of the protein-protein interaction by which CRP is able to recruit RNAP and up-regulate transcription.  

Partially mutagenized sequences have also been used extensively for ``deep mutational scanning'' experiments on proteins, starting with \cite{Fowler:2010gt,Hietpas:2011bp}. In this context, selection experiments on partially mutagenized proteins allow one to identify protein domains critical for folding and function. A variety of deep mutational scanning experiments are described in \cite{Fowler:2014gq}.

\section{Inference using likelihood}

\begin{figure}[t] 
   \centering
   \includegraphics{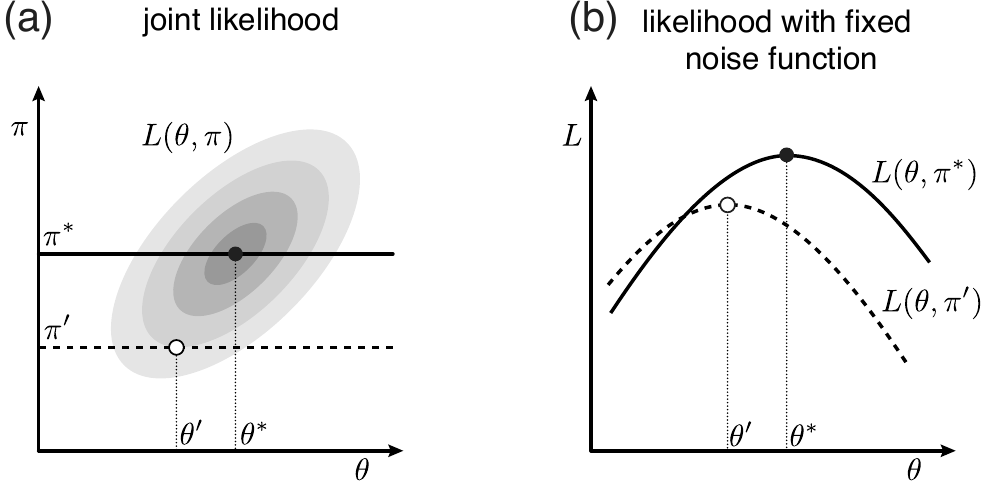} 
   \caption{Schematic illustration of how likelihood $L(\theta,\pi)$ depends on the model $\theta$ and the noise function $\pi$ in the $N \to \infty$ limit. (a,b) $L$ will typically have a correlated dependence on $\theta$ and $\pi$. If $\pi$ is set equal to the correct noise function $\pi^*$, then $L$ will be maximized by the correct model $\theta^*$. However, if $\pi$ is set to an incorrect noise function $\pi'$, $L$ will typically attain a maximum at an incorrect $\theta'$.}
   \label{fig:likelihood}
\end{figure}

The inference of quantitative sequence-function relationships from massively parallel experiments can be phrased as follows. Data consists of a large number of sequences $\set{S_n}_{n=1}^N$, each sequence $S$ having a corresponding measurement $M$. Due to experimental noise, repeated measurements of the same sequence $S$ can yield different values for $M$. Our experiment therefore has the following probabilistic form form:
\begin{diagram}[LaTeXeqno]
\underset{\rm sequence}{S} & \rTo^{\stackrel{\rm experiment}{p(M|S)}} & \underset{\rm measurement}{M}. 
\end{diagram}
If we assume that the measurements for each sequence are independent, and if we have an explicit parametric form for $p(M|S)$, then we can learn the values of the parameters by maximizing the per-datum log likelihood,
\bea
L = \frac{1}{N} \sum_{n=1}^N \log p(M_n | S_n).
\eea
In what follows we will refer to the quantity $L$ simply as the ``likelihood.'' 

In regression problems such as this, one introduces an additional layer of structure. Specifically, we assume the measurement $M$ of each sequence $S$ is a noisy readout of some underlying activity $R$ that is a deterministic function of that sequence.  We call the function relating $R$ to $S$ the ``activity model'' and denote it using $\theta(S)$. This activity model is ultimately what we want to understand. The specific way the activity $R$ is read out by measurements $M$ is then specified by a conditional probability distribution, $\pi(M|R)$, which we call the ``noise function.''\footnote{We use the term ``noise function'' in order to be consistent with the terminology of \cite{Kinney:2014ge} and to avoid deviating too much from the more standard terms ``noise model'' and ``error model'' used in the statistics and machine learning literature. We emphasize, however, that $\pi$ defines much more than just the characteristics of experimental noise; $\pi$ entirely specifies the relationship between measurements $M$ and the underlying activity $R$. Were it not for prior terminology, the term ``measurement function'' might be preferable to ``noise function.''} Our experiment is thus represented by the Markov chain
\begin{diagram}[LaTeXeqno]
\underset{\rm sequence}{S} & \rTo^{\stackrel{\rm model}{\theta(S)}} & \underset{\rm activity}{R} & \rTo^{\stackrel{\rm noise~function}{\pi(M|R)}} & \underset{\rm measurement}{M}. \label{eq:srm}
\end{diagram}
The corresponding likelihood is
\begin{eqnarray}
L(\theta,\pi) = \frac{1}{N} \sum_{n=1}^N \log \pi(M_n | \theta(S_n)). \label{eq:likelihood}
\end{eqnarray}
The model we adopt for our experiment therefore has two components: $\theta$, which describes the sequence-function relationship of interest, and $\pi$, which we do not really care about. 

Standard statistical regression requires that the noise function $\pi$ be specified up-front. $\pi$ can be learned either by performing separate calibration experiments, or by assuming a functional form based on an educated guess. This can be problematic, however. Consider inference in the large data limit, $N \to \infty$, which is illustrated in Fig.\ \ref{fig:likelihood}. Likelihood is determined by both the model $\theta$ and the noise function $\pi$ (Fig.\ \ref{fig:likelihood}a). If we know the correct noise function $\pi^*$ exactly, then maximizing $L(\theta,\pi^*)$ over $\theta$ is guaranteed to recover the correct model $\theta^*$. However, if we assume an incorrect noise function $\pi'$, maximizing likelihood will typically recover an incorrect model $\theta'$ (\fig{fig:likelihood}{b}). 

\section{Inference using mutual information}

Information theory provides an alternative inference approach. Suppose we hypothesize a specific model $\theta$, which gives predictions $R$. Denote the true model $\theta^*$ and the corresponding true activity $R^*$. The dependence between $S$, $M$, $R^*$, and $R$ will then form a Markov chain,
\begin{diagram}[LaTeXeqno]
R & \lTo^{\theta} & S & \rTo^{\theta^*} & R^* & \rTo^{\pi} & M. \label{eq:srm}
\end{diagram}
From the simple fact that $M$ depends on $R$ only through the value of $R^*$, any dependence measure $\mathcal{D}$ that satisfies the Data Processing Inequality (DPI) \cite{Cover:1991ti} must satisfy
\bea
\mathcal{D}[R;M] \leq \mathcal{D}[R^*;M].
\eea
Therefore, in the set of possible models $\theta$, the true model is guaranteed to globally maximize the objective function $\mathcal{D}(\theta) \equiv \mathcal{D}[R;M]$. 

One particularly relevant dependence measure that satisfies DPI is mutual information, a quantity that plays a fundamental role in information theory \cite{Cover:1991ti}.\footnote{See \cite{Kinney:2014vn} for an extended discussion of mutual information as a measure of statistical association.} For the massively parallel experiments such as those in Fig.\ 2, $R$ is continuous and $M$ is discrete. In these cases, mutual information is given by
\begin{eqnarray}
I(\theta) = I[R;M] = \sum_M \int dR\,  p(R,M) \log \frac{p(R,M)}{p(R)p(M)}, \label{eq:mi}
\end{eqnarray}
where $p(M,R)$ is the joint distribution of activity predictions and measurements resulting from the model $\theta$. If one is able to estimate $p(M,R)$ from a finite sample of data, mutual information can be used as an objective function for determining $\theta$ without assuming any noise function $\pi$. 

It should be noted that there are multiple dependence measures $\mathcal{D}$ that satisfy DPI. One might wonder whether maximizing multiple different dependence measures would improve on the optimization of mutual information alone. The answer is not so simple. In \cite{Kinney:2014ge} it was shown that if the correct model $\theta^*$ is within the space of models under consideration, then, in the large data limit, maximizing mutual information is equivalent to simultaneously maximizing every dependence measure that satisfies DPI. On the other hand, one rarely has any assurance that the correct model $\theta^*$ is within the space of parameterized models one is considering. In this case, considering different DPI-satisfying measures might provide a test for whether $\theta^*$ is noticeably outside the space of parameterized models. To our knowledge, this potential approach to the model selection problem has yet to be demonstrated. 

\section{Relationship between likelihood and mutual information}

A third inference approach is to admit that we do not know the noise function $\pi$ \emph{a priori}, and to fit \emph{both} $\theta$ and $\pi$ simultaneously by maximizing $L(\theta, \pi)$ over this pair. It is easy to see why this makes sense: the division of the inference problem into first measuring $\pi$, then learning $\theta$ using that inferred $\pi$, is somewhat artificial. The process that maps $S$ to $M$ is determined by both $\theta$ and $\pi$ and thus, from a probabilistic point of view, it makes sense to maximize likelihood over both of these quantities simultaneously. 

We now show that, in the large $N$ limit, maximizing likelihood over both $\theta$ and $\pi$ is equivalent to maximizing the mutual information between model predictions and measurements. Here we follow the argument given in \cite{Kinney:2014ge}. In the large $N$ limit, likelihood can be written
\begin{eqnarray}
L(\theta,\pi) &=& \sum_M \int dR\, p(R,M) \log \pi(M|R) \label{eq:L} \\
&=& I(\theta) - D(\theta,\pi) - H[M],  \label{eq:decomposition}
\end{eqnarray}
where 
\begin{eqnarray}
D(\theta, \pi) = \sum_M \int dR\, p(R,M) \log \frac{p(M|R)}{\pi(M|R)},
\end{eqnarray}
is the Kullback-Leibler divergence between the assumed noise function $\pi$ and the observed noise function $p(M | R)$, and $H[M] = - \sum_M p(M) \log p(M)$ is the entropy of the measurements, which does not depend on $\theta$. To maximize $L(\theta, \pi)$ it therefore suffices to maximize $I(\theta)$ over $\theta$ alone, then to set the noise function $\pi(M|R)$ equal to the empirical noise function $p(M|R)$, which causes $D(\theta,\pi)$ to vanish. 

Thus, when we are uncertain about the noise function $\pi$, we need not despair. We can, if we like, simply learn $\pi$ at the same time that we learn $\theta$. We need not explicitly model $\pi$ in order to do this; it suffices instead to maximize the mutual information $I(\theta)$ over $\theta$ alone. 

The connection between mutual information and likelihood can further be seen in a quantity called the ``noise-averaged'' likelihood. This quantity was first described for the analysis of microarray data \cite{Kinney:2007dh}; see also \cite{Kinney:2014ge}. The central idea is to put an explicit prior on the space of possible noise functions, then compute likelihood after marginalizing over these noise functions. Explicitly, the per-datum log noise-averaged likelihood $L_\mathrm{na}(\theta)$ is related to $L(\theta,\pi)$ via
\bea
e^{N L_\mathrm{na}(\theta)} &=& \int d\pi\, p(\pi)\, e^{N L(\theta, \pi)}.
\eea
We will refer to $L_\mathrm{na}$ simply as ``noise-averaged likelihood'' in what follows. 

Under fairly general conditions, one finds that noise-averaged likelihood is related to mutual information via
\begin{eqnarray}
L_\mathrm{na}(\theta) &=& I(\theta) - \Delta(\theta) - H[M]. \label{eq:marginal_log_likelihood}
\end{eqnarray}
Here, the effect of the noise function prior $p(\pi)$ is absorbed entirely by the term $\Delta(\theta)$. Under very weak assumptions, $\Delta(\theta)$ vanishes in the $N \to \infty$ limit and thus $p(\pi)$ becomes irrelevant for the inference problem \cite{Kinney:2007dh,Kinney:2014ge}.

\section{Diffeomorphic modes}

Mutual information has a mathematical property that is important to account for when using it as an objective function: the mutual information between any two variables is unchanged by an invertible transformation of either variable. So if a change in model parameters, $\theta \to \theta'$, results in changes in model predictions $R \to R'$ that preserves the rank order of these predictions, then
\bea
I(\theta) = I[M;R] = I[M;R'] = I(\theta'),
\eea
and $\theta$ and $\theta'$ are judged to be equally valid. 

\begin{figure}[t] 
   \centering
   \includegraphics{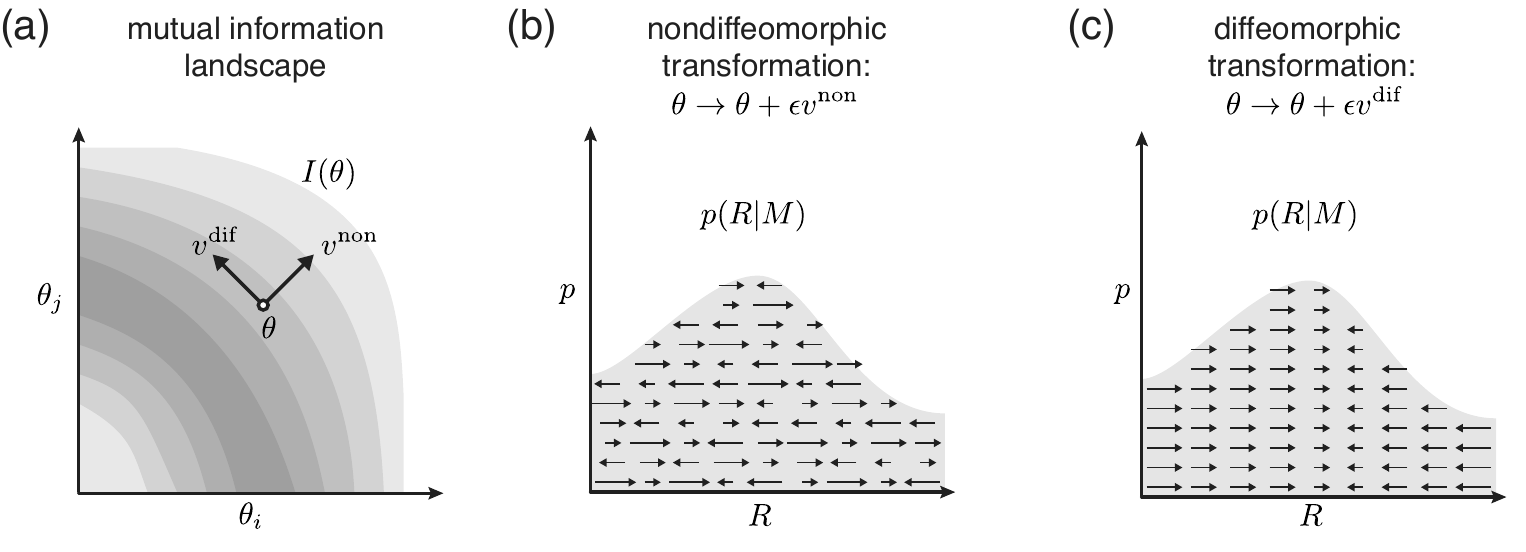} 
   \caption{Illustration of diffeomorphic and nondiffeomorphic modes. (a) A diffeomorphic mode $v^\mathrm{dif}$ at a point $\theta$ in parameter space is a vector that will (regardless of the underlying data) be tangent to a level curve of $I(\theta)$. All other vectors (e.g., $v^\mathrm{non}$) correspond to nondiffeomorphic modes. (b) Moving $\theta$ along a nondiffeomorphic mode results in a sort of ``diffusion'' in which the $R$ values assigned to different sequences change rank order. Here, the probability distribution $p(R|M)$ is illustrated (for fixed $M$) in gray. The motion of individual $R$ values upon such a change in $\theta$ are indicated by arrows. (c) Changing $\theta$ along a diffeomorphic mode, however, results in a ``flow'' of $R$ values that maintains their rank order.}
   \label{fig:particle_motions}
\end{figure}

By using mutual information as an objective function, we are therefore unable to constrain any parameters of $\theta$ that, if changed, produce invertible transformations of model predictions. Such parameters are called ``diffeomorphic parameters'' or ``diffeomorphic modes'' \cite{Kinney:2014ge}. The distinction between diffeomorphic modes and nondiffeomorphic modes is illustrated in \fig{fig:particle_motions}. 

\subsection{Criterion for diffeomorphic modes}

Following \cite{Kinney:2014ge}, we now derive a criterion that can be used to identify all of the diffeomorphic modes of a model $\theta$.\footnote{Here, as throughout this paper, we restrict our attention to situations in which $R$ is a scalar. The case of vector-valued model predictions $R$ is worked out in \cite{Kinney:2014ge}.} Consider an infinitesimal change in model parameters $\theta \to \theta + d \theta$, where the components of $d\theta$ are specified by
\bea
d \theta_i = \epsilon v_i
\eea
for small epsilon and for some vector $v_i$ in $\theta$-space. This change in $\theta$ will produce a corresponding change in model predictions $R \to R + dR$, where
\bea
dR =  \epsilon \sum_i v_i \frac{\partial R}{\partial \theta_i}. \label{eq:change_in_R}
\eea
In general, the derivative $\partial R / \partial \theta_i$ can have arbitrary dependence on the underlying sequence $S$. This transformation will preserve the rank order of $R$-values only if $dR$ is the same for all sequences having the same value of $R$. The change $dR$ must therefore be a function of $R$ and have no other dependence on $S$. A diffeomorphic mode is a vector field $v^\mathrm{dif}(\theta)$ that has this property at all points in parameter space. Specifically, a vector field $v^\mathrm{dif}(\theta)$ is a diffeomorphic mode if and only if there is a function $h(R,\theta)$ such that
\begin{eqnarray}
\sum_i v^\mathrm{dif}_i(\theta) \frac{\partial R}{\partial \theta_i}  = h(R,\theta).\label{eq:diff_mode_def}
\end{eqnarray}

\subsection{Diffeomorphic modes of linear models}

As a simple example, consider a situation in which each sequence $S$ is a $D$-dimensional vector and $R$ is an affine function of $S$, i.e.
\bea
R = \theta_0 + \sum_{i=1}^D \theta_i S_i, \label{eq:linear_model}
\eea 
for model parameters $\theta = \set{\theta_0, \theta_1, \ldots, \theta_D}$. The criterion in \eq{eq:diff_mode_def} then gives
\bea
v^\mathrm{dif}_0(\theta) + \sum_{i=1}^D v^\mathrm{dif}_i(\theta) S_i = h(R,\theta).
\eea
Because the left hand side is linear in $S$ and $R$ is linear in $S$, the function $h(R,\theta)$ must be linear in $R$. Thus, $h$ must have the form
\bea
h(R,\theta) = a(\theta) + b(\theta) R \label{eq:linear_mode}
\eea
for some functions $a(\theta)$ and $b(\theta)$. The corresponding diffeomorphic mode is 
\bea
v^\mathrm{dif}_i(\theta) = \left\{ 
\begin{array}{ccl}
a(\theta) &~~ & i = 0 \\
b(\theta) \theta_i & &i = 1, 2, \ldots, D 
\end{array}
\right. ,
\eea
which has two degrees of freedom. Specifically, the $a$ component of $v^\mathrm{dif}$ corresponds to adding a constant to $R$ while the $b$ component corresponds to multiplying $R$ by a constant. 

Note that if we had instead chosen $R = \sum_{i=1}^D \theta_i S_i$, i.e. left out the constant component $\theta_0$, then there would be only one diffeomorphic mode, corresponding to multiplication of $R$ by a constant. This fact will be used when we analyze the Gaussian selection model in Section 8.

\subsection{Diffeomorphic modes of a biophysical model of transcriptional regulation}

Diffeomorphic modes can become less trivial in more complicated situations. Consider the biophysical model of transcriptional regulation by the \emph{E. coli} \emph{lac} promoter (\fig{fig:transcription}{}). This model was fit to Sort-Seq data in \cite{Kinney:2010tn}. The form of this model is as follows. Let $S$ denote a $4 \times D$ matrix representing a DNA sequence of length $D$ and having elements
\bea
S_{bl} = \left\{ \begin{array}{ccl} 1 & ~& \mathrm{if~base~}b\mathrm{~occurs~at~position~}l \\ 0 &~& \mathrm{otherwise} \end{array} \right.
\eea
where $b \in \set{A,C,G,T}$ and $l = 1, 2, \ldots D$. The binding energy $Q$ of CRP to DNA was modeled in \cite{Kinney:2010tn} as an ``energy matrix'': each position in the DNA sequence was assumed to contribute additively to the overall energy. Specifically, 
\bea
Q = \sum_{b,l} \theta_{Q}^{bl} S_{bl} + \theta^0_Q,
\eea
where $\theta_Q = \set{\theta^0_Q, \theta^{bl}_Q}$ are the parameters of this energy matrix. Similarly, the binding energy $P$ of RNAP to DNA was modeled as 
\bea
P = \sum_{b,l} \theta_{P}^{bl} S_{bl} + \theta^0_P.
\eea
Both energies were taken to be in thermal units ($k_B T$). The rate of transcription $R$ resulting from these binding energies was assumed to be proportional to the occupancy of RNAP at its binding site. This is given by
\begin{eqnarray}
R = R_\mathrm{max}\, \frac{e^{-P} + e^{-P - Q - \gamma}}{1 + e^{-Q} + e^{-P} + e^{-P - Q - \gamma}}, \label{eq:kinney_model}
\end{eqnarray}
where $\gamma$ is the interaction energy between CRP and RNAP (again in units of $k_B T$). 

Because the binding sites for CRP and RNAP do not overlap, one can learn the parameters $\theta_Q$ and $\theta_P$ from data separately by independently maximizing $I[Q;M]$ and $I[P;M]$. Doing this, however, leaves undetermined the overall scale of each energy matrix as well as the chemical potentials $\theta_P^0$ and $\theta_Q^0$. The reason is that the energy scale and chemical potential are diffeomorphic modes of energy matrix models and therefore cannot be inferred by maximizing mutual information. 

However, if $Q$ and $P$ are inferred together by maximizing $I[R;M]$ instead, one is now able to learn both energy matrices with a physically meaningful energy scale. The chemical potential of CRP, $\theta_Q^0$, is also determined. The only parameters left unspecified are the chemical potential of RNA polymerase, $\theta_P^0$, and the maximal transcription rate $R_\mathrm{max}$. The reason for this is that in the formula for $R$ in \eq{eq:kinney_model} the energies $P$ and $Q$ combine in a nonlinear way.  This nonlinearity eliminates three of the four diffeomorphic modes of $P$ and $Q$.\footnote{The one additional diffeomorphic mode is created by the introduction of the parameter $R_\mathrm{max}$.} See \cite{Kinney:2014ge} for the derivation of this result. 

\subsection{Dual modes of the noise function} \label{sec:dualmodes}

\begin{figure}[t] 
   \centering
   \includegraphics{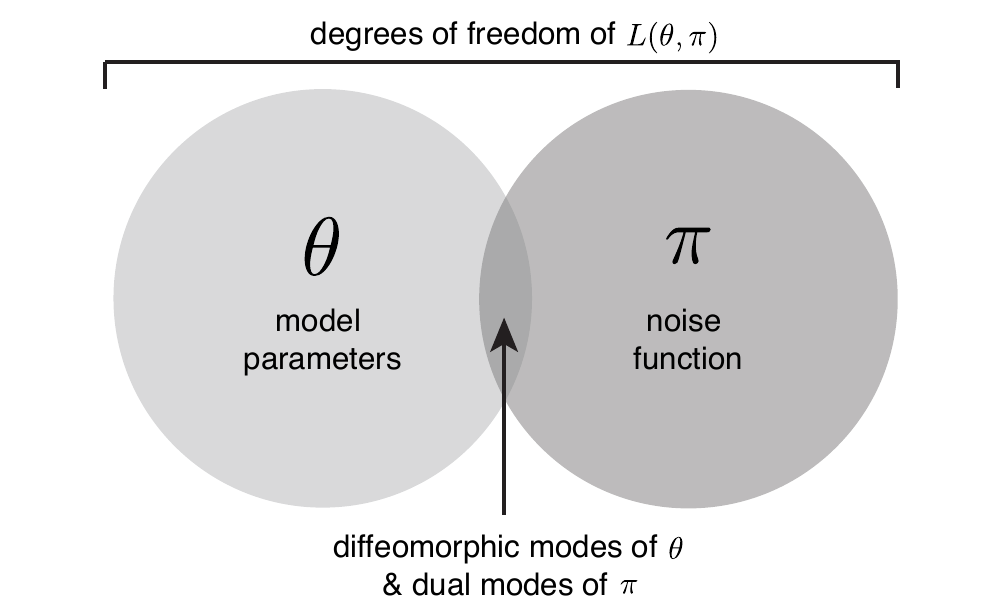} 
   \caption{Venn diagram illustrating the degrees of freedom of the likelihood $L(\theta,\pi)$ considered over all possible data sets $\set{S_n, M_n}$. Altering the model parameters $\theta$ will typically change $L(\theta,\pi)$ in a way that cannot be recapitulated by changes in the noise function $\pi$. Similarly, changes in $\pi$ cannot typically be imitated  by changes in $\theta$. However, diffeomorphic transformations of $\theta$ will affect $L(\theta,\pi)$ in the exact same way that dual transformation of $\pi$ will. The diffeomorphic modes of $\theta$ and the dual modes of $\pi$ can therefore be thought of as lying within the intersection of $\theta$ and $\pi$.}
   \label{fig:parameters}
\end{figure}

Diffeomorphic transformations of model parameters can be thought of as being equivalent to certain transformations of the noise function $\pi$. Consider the transformation of model parameters 
\bea
\theta_i \to \theta_i' = \theta_i + \epsilon v_i, \label{eq:theta_transformation}
\eea
where $\epsilon$ is an infinitesimal number and $v_i$ is a vector in $\theta$-space.\footnote{For the sake of clarity we suppress the $\theta$-dependence of $v^\mathrm{dif}$, $\tilde{v}^\mathrm{dif}$, and $h(R)$ in what follows.} For any sequence $S$, this transformation induces a transformation of the model prediction
\bea
R \to R' &=& R + \epsilon \sum_i  v_i \frac{\partial R}{\partial \theta_i }. \label{eq:ab}  
\eea
To see the effect this transformation has on likelihood, we rewrite \eq{eq:likelihood} as,
\bea
L(\theta, \pi) = \braket{\log \pi(M|R)}_\data,
\eea
where $\braket{\cdot}_\data$ indicates an average taken over the measurements $M_n$ and predictions $R_n$ for all of the sequences $S_n$ in the data set. The change in likelihood resulting from \eq{eq:ab} is therefore given by
\bea
L(\theta',\pi) &=& L(\theta, \pi) + \epsilon \braket{\frac{\partial \log \pi(M|R)}{\partial R}\sum_i \frac{\partial R}{\partial \theta_i } v_i }_\data. \label{eq:likelihood_transformed_theta}
\eea

Now suppose that there is a noise function $\pi'$ that has an equivalent effect on likelihood, i.e.,
\bea
L(\theta',\pi) = L(\theta,\pi') + O(\epsilon^2),
\eea
for all possible data sets $\set{S_n, M_n}$. We say that this transformation of the noise function $\pi \to \pi'$ is  ``dual'' to the transformation $\theta \to \theta'$ of model parameters. The transformed noise function will necessarily have the form
\bea
\log \pi'(M|R) =  \log \pi(M|R) +  \epsilon \tilde{v}(M,R)  \label{eq:pi_transformation}
\eea
for some function $\tilde{v}(M,R)$. To determine $\tilde{v}$ we consider the transformation of likelihood induced by $\pi \to \pi'$: 
\bea
L(\theta,\pi') &=&  L(\theta,\pi) + \epsilon \braket{\tilde{v}(M,R)}_\data.\label{eq:likelihood_transformed_pi}
\eea
Comparing \eq{eq:likelihood_transformed_theta} and \eq{eq:likelihood_transformed_pi}, we see that $\pi \to \pi'$ will be dual to $\theta \to \theta'$ for all possible data sets if and only if
\bea
\frac{\partial \log \pi(M|R)}{\partial R} \sum_i \frac{\partial R}{\partial \theta_i } v_i = \tilde{v}(M,R) \label{eq:duality_condition}
\eea
for all sequences $S$.

For general choice of vector $v$, no function $\tilde{v}$ will exist that satisfies \eq{eq:duality_condition}. The reason is that $\partial R/\partial \theta_i$ will typically depend on the sequence $S$ independently of the value of $R$. In other words, for a fixed value of $M$ and $R$, the left hand side of \eq{eq:duality_condition} will retain a dependence on $S$. The right hand side, however, cannot have such a dependence. The converse is also true: for general choice of the function $\tilde{v}$, no vector $v$ will exist such that \eq{eq:duality_condition} is satisfied for all sequences. This is evident from the simple fact that $v$ is a finite dimensional vector while $\tilde{v}$ is a function of the continuous quantity $R$ and therefore has an infinite number of degrees of freedom.

In fact, \eq{eq:duality_condition} will have a solution if and only if
\bea
\sum_i \frac{\partial R}{\partial \theta_i } v^\mathrm{dif}_i = h(R)
\eea
for some function $h$. Here we have added the superscript ``dif'' because this is precisely the definition of a diffeomorphic mode given in \eq{eq:diff_mode_def}. In this case, the function $\tilde{v}^\mathrm{dif}$ dual to this diffeomorphic mode $v^\mathrm{dif}$ is seen to be 
\bea
\tilde{v}^\mathrm{dif}(M,R) = \frac{\partial \log \pi(M|R)}{\partial R} h(R).
\eea

These findings are summarized by the Venn diagram in \fig{fig:parameters}{}. Arbitrary transformations of the model parameters $\theta$ will alter likelihood in a way that cannot be imitated by any change to the noise function $\pi$. The reverse is also true: most changes to $\pi$ cannot be imitated by a corresponding change in $\theta$. However, a subset of transformations of $\theta$ are equivalent to corresponding dual transformations of $\pi$. These transformations are precisely the diffeomorphic transformations of $\theta$. This partial duality between $\theta$ and $\pi$ has a simple interpretation: the choice of how we parse an experiment into an activity model $\theta$ and a noise function $\pi$ is not unique. The ambiguity in this choice is parameterized by the diffeomorphic modes of $\theta$ and the dual modes of $\pi$. 

\section{Error bars from likelihood, mutual information, and noise-averaged likelihood}

\begin{figure}[t] 
   \centering
   \includegraphics{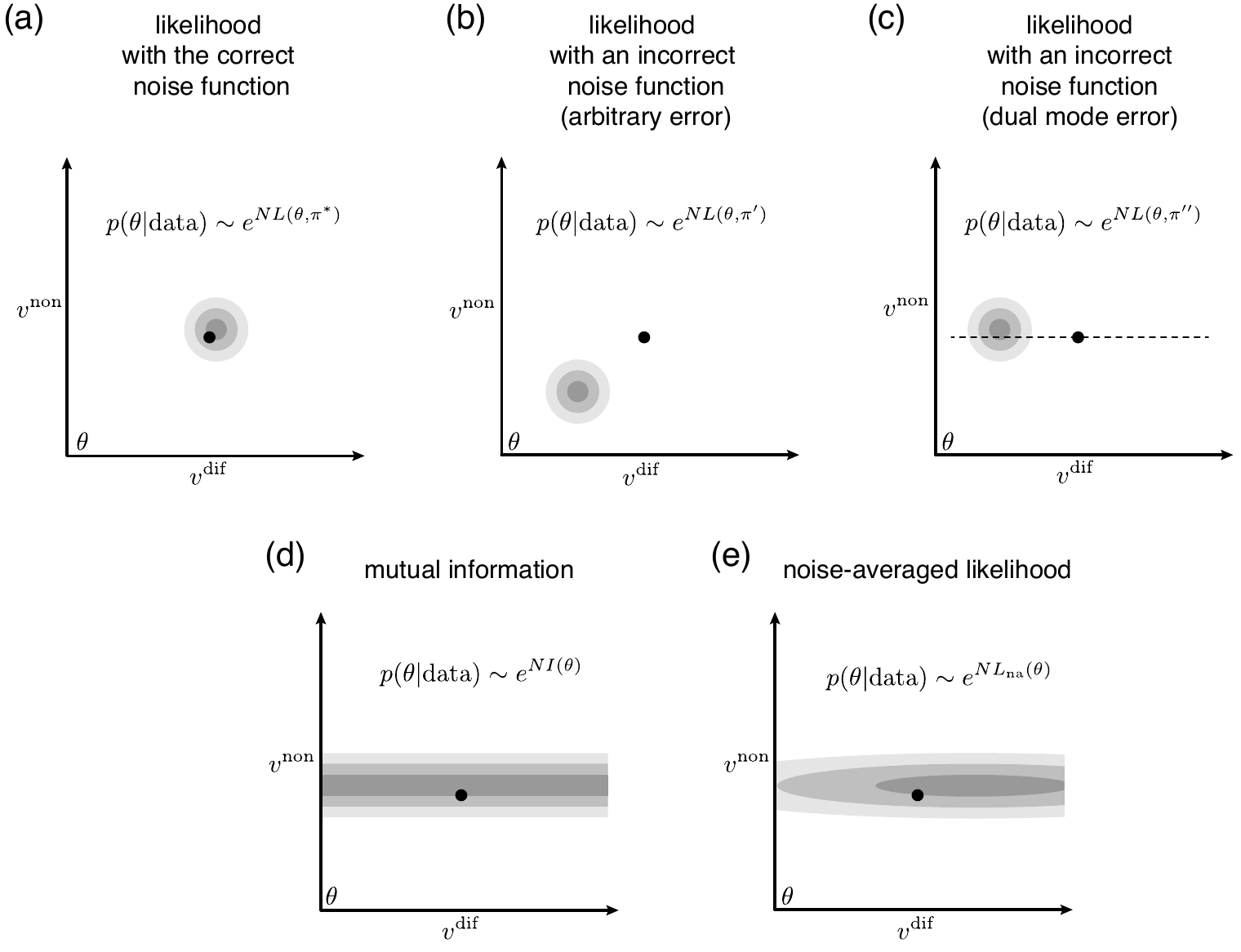} 
   \caption{Posterior distributions on model parameters resulting from various objective functions. Each panel schematically illustrates the posterior distribution $p(\theta | \data)$ (gray shaded area) as it relates to the correct model $\theta^*$ (dot) along both diffeomorphic (abscissa) and nondiffeomorphic (ordinate) directions in parameter space. (a) Likelihood with the correct noise function $\pi^*$ leads to a posterior distribution consistent with $\theta^*$ in all parameters. (b) Likelihood with a noise function $\pi'$ that differs arbitrarily from $\pi^*$ will, in general, lead to a posterior distribution that is inconsistent with $\theta^*$ along both diffeomorphic and nondiffeomorphic modes. (c) Likelihood with a noise function $\pi''$ that differs from $\pi^*$ only along a dual mode $\tilde{v}^\mathrm{dif}$ leads to a posterior that is inconsistent with $\theta^*$ along the diffeomorphic mode $v^\mathrm{dif}$ (parallel to dashed line), but consistent with $\theta^*$ in all other directions  (perpendicular to dashed line). (d) Using mutual information give a posterior that is consistent with $\theta^*$; this posterior places constraints similar to likelihood along non-diffeomorphic modes but places no constraints whatsoever along diffeomorphic modes. (e) Using noise-averaged likelihood results in a posterior distribution similar to mutual information but with weak constraints on diffeomorphic modes resulting from the noise function prior $p(\pi)$.}
   \label{fig:hessians}
\end{figure}

We now consider the consequences of performing inference using various objective functions at large but finite $N$. Specifically, we discuss the optimal parameters and corresponding error bars that are found by sampling $\theta$ from posterior distributions of the form 
\bea
p(\theta | \data) \sim e^{N F(\theta)} \label{eq:sampling_from_F}
\eea
for the following choices of the objective function $F(\theta)$:
\begin{enumerate}[(a)]
\item $F(\theta) = L(\theta,\pi^*)$ is likelihood computed using the correct noise function $\pi^*$.
\item $F(\theta) = L(\theta, \pi')$ where $\pi'$ differs from $\pi^*$ by a small but arbitrary error.
\item $F(\theta) = L(\theta, \pi'')$ where $\pi''$ differs from $\pi^*$ by a small amount along a dual mode. 
\item $F(\theta) = I(\theta)$ is the mutual information between measurements and model predictions.
\item $F(\theta) = L_\mathrm{na}(\theta)$ is the noise-averaged likelihood. 
\end{enumerate} 

To streamline notation, we will use $\braket{\cdot}$ to denote averages computed in multiple different contexts. In each case, the appropriate context will be specified by a subscript. As above $\braket{\cdot}_\data$ will denote averaging over a specific data set $\set{S_n,M_n}_{n=1}^N$. $\braket{\cdot}_\mathrm{real}$ will indicate averaging over an infinite number of data set realizations. $\braket{\cdot}_S$, $\braket{\cdot}_{S,M}$, $\braket{\cdot}_{S|R}$, and $\braket{\cdot}_{S|R,M}$ will respectively denote averages over the distributions $p(S)$, $p(S,M)$, $p(S|R)$, and $p(S|R,M)$, the empirical distributions obtained in the infinite data limit. $\braket{\cdot}_\theta$ will indicate an average computed over parameter values $\theta$ sampled from the posterior distribution $p(\theta|\data)$. Subscripts on $\cov(\cdot)$ or $\var(\cdot)$ should be interpreted analogously. 

\subsection{Likelihood}

Consider \eq{eq:sampling_from_F} with $F(\theta) = L(\theta,\pi^*)$ at large but finite $N$. The posterior distribution $p(\theta | \data)$ will, in general, be maximized at some choice of parameters $\theta^o$ that deviates randomly from the correct parameters $\theta^*$. At large $N$, $p(\theta|\data)$ will become sharply peaked about $\theta^o$ with a peak width governed by the Hessian of likelihood; specifically
\bea
\cov_\theta(\theta_i - \theta^o_i, \theta_j - \theta^o_j) = - \frac{H_{ij}^{-1}}{N},
\eea
where 
\bea
H_{ij} = \left. \frac{\partial^2 L(\theta,\pi^*)}{\partial \theta_i \partial \theta_j} \right|_{\theta^*},  \label{eq:likelihood_hessian}
\eea
is the Hessian of the likelihood.  It is also readily shown (see Appendix A) that this peak width is consistent with the correct parameters $\theta^*$, in the sense that
\bea
\cov_\mathrm{real}(\theta_i^* - \theta_i^o,\theta_j^* - \theta_j^o) = \cov_\theta(\theta_i - \theta^o_i, \theta_j - \theta^o_j). \label{eq:theta_o_cov}
\eea

In Appendix A we show that the Hessian of likelihood, \eq{eq:likelihood_hessian}, is given by
\bea
H_{ij} &=& - \int dR\, p(R) J(R) \left. \braket{\frac{\partial R}{\partial \theta_i} \frac{\partial R}{\partial \theta_j} }_{S|R} \right|_{\theta^*}, \label{eq:likelihood_hessian_formula} 
\eea
where  
\bea
J(R) = 
\sum_M \pi^*(M|R) \left[ \frac{\partial \log \pi^*(M|R)}{\partial R} \right]^2 
= - \sum_M \pi^*(M|R) \frac{\partial^2 \log \pi^*(M|R)}{\partial R^2}
\label{eq:fisher_information}
\eea
is the Fisher information of the noise function $\pi^*$. This Fisher information is a nonnegative measure of how sensitive our experiment is in the vicinity of $R$.\footnote{In what follows we assume that $J(R) > 0$ almost everywhere. This just reflects the assumption that our experiment actually does convey information about $R$ through the measurements $M$ it provides.} We thus see that, as long as the set of vectors $\partial R/\partial \theta_i$ spans all directions in parameter space, the Hessian matrix $H_{ij}$ will be nonsingular. Using $F(\theta) = L(\theta,\pi^*)$ will therefore put constraints on all directions in parameters space, and these constraints will shrink with increasing data as $N^{-1/2}$. This situation is illustrated in Fig.\ 7a. 

Now consider what happens if instead we use a noise function $\pi'$ that deviates from $\pi^*$ in a small but arbitrary way. Specifically, let
\bea
\log \pi'(M|R) = \log \pi^*(M|R) + \epsilon f(M,R) \label{eq:noise_function_deviation}
\eea
for some function $f(M,R)$ and small parameter $\epsilon$. It is readily shown (see Appendix A) that the maximum likelihood parameters $\theta'$ will deviate from $\theta^*$ by an amount
\bea
\braket{\theta'_i - \theta_i^*}_\mathrm{real} = - \epsilon \sum_j H^{-1}_{ij} w_j,~~~~\mathrm{where}~~~~w_j = \left. \braket{ \frac{\partial f}{\partial R} \frac{\partial R}{\partial \theta_j} }_S \right|_{\theta^*}. \label{eq:deviation}
\eea
This expected deviation does not depend on $N$ and will therefore not shrink to zero in the large $N$ limit. Indeed, for any choice of $\epsilon > 0$, there will always be an $N$ large enough such that this bias in $\theta'$ dominates over the uncertainty due to finite sampling. 

Is there any restriction on the types of biases in $\theta'$ that can be produced by the choice of incorrect noise function $\pi'$? In general, no. Because the Hessian matrix $H$ is nonsingular, one can always find a vector $w$ such that the deviation of $\theta'$ from $\theta^*$ in \eq{eq:deviation} points in any chosen direction of $\theta$-space. As long as the functions
\bea
g_i(R) = \left. \braket{\frac{\partial R}{\partial \theta_i}}_{S|R} \right|_{\theta^*}
\eea
are linearly independent for different indices $i$, a function $f$ can always be found that generates the vector $w$. 

We therefore see that arbitrary errors in the noise function will bias the inference of model parameters in arbitrary directions. This fact presents a major concern for standard likelihood-based inference: if you assume an incorrect noise function $\pi$, the parameters $\theta$ that you then infer will, in general, be biased in an unpredictable way. Moreover, the magnitude of this bias will be directly proportional to the magnitude of the error in the log of your assumed noise function. This problem is illustrated in Fig.\ 7b.

There is a case that deserves some additional consideration. Suppose we use a noise function $\pi''$ that differs from $\pi^*$ only along a dual mode $\tilde{v}^\mathrm{dif}$, i.e.,
\bea
\log \pi''(M|R) = \log \pi^*(M|R) + \epsilon \tilde{v}^\mathrm{dif}(M,R). \label{eq:dual_noise_function_deviation}
\eea
The maximum likelihood parameters $\theta''$ of $L(\theta,\pi'')$ will still deviate from $\theta^*$ by an amount that does not shrink to zero in the $N \to \infty$ limit. However, this bias in parameter values will be restricted to the diffeomorphic mode $v^\mathrm{dif}$ to which $\tilde{v}^\mathrm{dif}$ is dual, i.e.,
\bea
\braket{\theta''_i - \theta^*_i}_\mathrm{real} = - \epsilon v_i^\mathrm{dif}. \label{eq:theta_prime_prime}
\eea
This state of affairs ain't so bad since the incorrect noise function will lead to model parameters that are inaccurate only along modes that we already know we cannot learn from the data. This situation is illustrated in Fig.\ 7c; see Appendix A for the derivation of \eq{eq:theta_prime_prime}. 

\subsection{Mutual information}

The constraints on parameters imposed by using mutual information $I(\theta)$ as the objective function $F(\theta)$ in \eq{eq:sampling_from_F} are determined by the Hessian
\bea
K_{ij} = \left. \frac{\partial^2 I(\theta)}{\partial \theta_i \partial \theta_j} \right|_{\theta^*}.
\eea
Appendix B provides a detailed derivation of this Hessian, which after some computation is found to be given by
\bea
K_{ij} = - \int dR\, p(R) J(R) \left. \left[ 
	\braket{\frac{\partial R}{\partial \theta_i} \frac{\partial R}{\partial \theta_j}}_{S|R}
	- \braket{\frac{\partial R}{\partial \theta_i}}_{S|R} \braket{\frac{\partial R}{\partial \theta_j}}_{S|R}  
\right] \right|_{\theta^*}. \label{eq:mi_hessian_formula}
\eea
Comparing \eq{eq:mi_hessian_formula} and \eq{eq:likelihood_hessian_formula}, we see that for any vector $v$ in parameter space,
\bea
- \sum_{i,j} H_{ij} v_i v_j \ge - \sum_{i,j} K_{ij} v_i v_j \ge 0.
\eea
Likelihood is thus seen to constrain parameters in all directions at least as much as mutual information does. As expected, mutual information provides no constraint whatsoever in the direction of any diffeomorphic mode $v^\mathrm{dif}$ of the model, since
\bea
- \sum_{i,j} K_{ij} v_i^\mathrm{dif} v_j^\mathrm{dif} 
= \int dR\, p(R) J(R) \left. \left[ \braket{h^2(R)}_{S|R} - \braket{h(R)}_{S|R}^2 
\right] \right|_{\theta^*}= 0.
\eea
The converse is also true: if there is no constraint on parameters along $v$, then $v$ must be a diffeomorphic mode. This is because
\bea
- \sum_{i,j} K_{ij} v_i v_j = \int dR\, p(R)\, J(R)\, \left. \var \left( \sum_i v_i \frac{\partial R}{\partial \theta_i} \right)_{S|R} \right|_{\theta^*} .\label{eq:constraints_of_K}
\eea
Because $J(R)$ is positive almost everywhere, the right hand side of \eq{eq:constraints_of_K} can vanish only if $\sum_i v_i \frac{\partial R}{\partial \theta_i}$ does not differ between any two sequences that have the same $R$ value. There must therefore exist a function $h(R)$ such that $h(R) =  \sum_i v_i \frac{\partial R}{\partial \theta_i}$ for all sequences $S$. This is precisely the requirement in \eq{eq:diff_mode_def} that $v$ be a diffeomorphic mode. 

However, except along diffeomorphic modes, we can generally expect that the constraints provided by likelihood and by mutual information will be of the same magnitude. This situation is illustrated in Fig.\ 7d. Indeed, in the next section we will see an explicit example where all nondiffeomorphic constraints imposed by mutual information are the same as those imposed by likelihood. 

Before proceeding, we note that the relationship between the Hessians of likelihood and mutual information suggests an analogy to fluid mechanics. Consider a trajectory in parameter space given by $\theta_i(t) = t v_i$, where $t$ is time and $v$ is a velocity vector pointing in the direction of motion. This motion in parameter space will induce a motion in the prediction $R(t)$ that the model provides for every sequence $S$. The set of sequence $\set{S_n}$ thus presents us with a dynamic cloud of ``particles'' moving about in $R$-space. At $t = 0$, the quantity $\braket{\dot{R}^2}_{S|R}$ will be proportional to the average kinetic energy of particles at location $R$. The quantity $\braket{\dot{R}}^2_{S|R}$ will be proportional to the (per particle) kinetic energy of the bulk fluid element at $R$, a quantity that does not count energy due to thermal motion. In this way we see that $-\sum_{i,j} H_{ij} v_i v_j$ is a weighted tally of total kinetic energy, whereas $-\sum_{i,j} K_{ij} v_i v_j$ corresponds to a tally of internal thermal energy only, the kinetic energy of bulk motion having been subtracted out.  

\subsection{Noise-averaged likelihood}

Noise-averaged likelihood provides constraints in between those of likelihood, computed using the correct noise function, and those of mutual information. This is illustrated in Fig.\ 7e. Whereas mutual information provides no constraints whatsoever on the diffeomorphic modes of $\theta$, noise-averaged likelihood provides weak constraints in these directions. These soft constraints reflect the Hessian of $\Delta(\theta)$ in \eq{eq:marginal_log_likelihood}. The constraints along diffeomorphic modes, however, have an upper bound on how tight they can become in the $N \to \infty$ limit. This is because such constraints only reflect our prior $p(\pi)$ on the noise function, not the information we glean from data. 

\section{Worked example: Gaussian selection}

\begin{figure}[t] 
   \centering
   \includegraphics{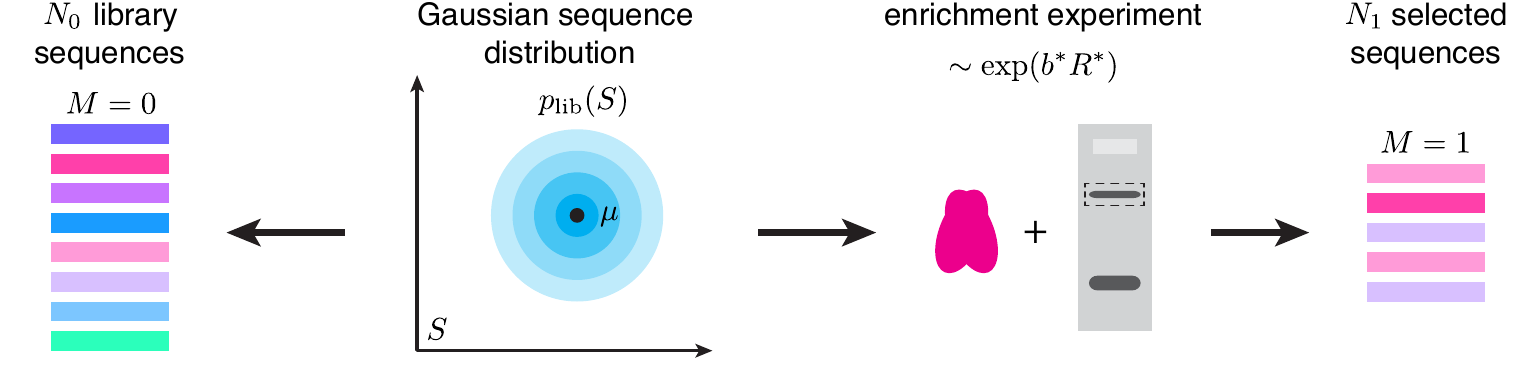} 
   \caption{ Illustration of the Gaussian selection model of a massively parallel experiment. Each assayed sequence in this model is a $D$-dimensional vector. The library (corresponding to bin $M=0$) consists of $N_0$ sequences $S$ drawn from a Gaussian distribution $p_\mathrm{lib}(S)$ that is centered on a specific sequence $\mu$.  Bin $M=1$ consists of $N_1$ sequences drawn from the distribution $p_\mathrm{lib}(S)$ then enriched by a factor of $\exp(b^* R^*)$ where $R^* = S^T \theta^*$. This enrichment procedure is analogous to selecting protein-bound DNA sequences where $b^* R^*$ is negative the binding energy. Calculations in the text are performed in the $N_0 \gg N_1$ limit.}
   \label{fig:gaussian_model}
\end{figure}

The above principles can be illustrated in the following analytically tractable model of a massively parallel experiment, which we call the ``Gaussian selection model.'' In this model, our experiment starts with a large library of ``DNA'' sequences $S$, each of which is actually a $D$-dimensional vector drawn from a Gaussian probability distribution\footnote{For the sake of simplicity we set the covariance matrix of this distribution equal to the identity matrix. The more general case of a non-identity covariance matrix yields the same basic results. Also, we note that, while approximating discrete DNA sequences by continuous vectors might seem crude, it is only the marginal distributions $p(R|M)$ that matter for the inference problem. Most of the quantities $R$ that one encounters in practice are computed by summing up contributions from a large number of different nucleotide positions. In such cases, the marginal distributions $p(R|M)$ will often be nearly continuous and virtually indistinguishable from the marginal distributions one might obtain from a Gaussian sequence library.}
\bea
p_\mathrm{lib}(S) = (2 \pi)^{-D/2} \exp \left( - \frac{|S - \mu|^2}{2} \right). \label{eq:library}
\eea
Here, $\mu$ is a $D$-dimensional vector defining the average sequence in the library. From this library we extract sequences into two bins, labeled $M=0$ and $M=1$. We fill the $M=0$ bin with sequences sampled indiscriminately from the library. The $M=1$ bin is filled with sequences sampled from this library with relative probability
\bea
\frac{p(M=1 | S)}{p(M=0|S)} = \exp( a^* + b^* R^*) \label{eq:selection_procedure}
\eea
where the activity $R^*$ is defined as the dot product of $S$ with  a $D$-dimensional vector $\theta^*$, i.e.,
\bea
R^* = S^T \theta^*.
\eea
We use $N_M$ to denote the number of sequences in each bin $M$, along with $N = N_0 + N_1$. 

All of our calculations are performed in the limit where $N_1$ is large but for which $N_0$ is far larger. More specifically, we assume that $\exp( a^* + b^* R^*) << 1$ everywhere that both $p(S|M=0)$ and $p(S|M=1)$ are significant. We use $\epsilon$ to denote the ratio
\bea
\epsilon \equiv \frac{p(M=1)}{p(M=0)} = \frac{N_1}{N_0}, \label{eq:epsilon_from_ps}
\eea
and all of our calculations are carried out only to first order in $\epsilon$. This model experiment is illustrated in \fig{fig:gaussian_model}{}.

Our goal is this: given the sampled sequences in the two bins, recover the parameters $\theta^*$ defining the sequence-function relationship for $R^*$. To do this, we adopt the following model for the sequence-dependent activity $R$:
\bea
R = S^T \theta, \label{eq:gaussian_activity_model}
\eea
where $\theta$ is the $D$-dimensional vector we wish to infer. From the arguments above and in \cite{Kinney:2014ge}, it is readily seen that the magnitude of $\theta$, i.e. $|\theta|$, is the only diffeomorphic mode of the model: changing this parameter rescales $R$, which preserves rank order. 

\subsection{Bin-specific distributions}
 
We can readily calculate the conditional sequence distribution $p(S|M)$ for each bin $M$, as well as the conditional distribution $p(R|M)$ of model predictions. Because the sequences sampled for bin 0 are indiscriminately drawn from $p_{\mathrm{lib}}$, we have 
\bea
p(S | M=0) &=& p_\mathrm{lib}(S) =  (2 \pi)^{-D/2} \exp \left( - \frac{|S - \mu|^2}{2} \right). \label{eq:null_library} 
\eea
The selected distribution of sequences is found to be
\bea
p(S | M=1) &=& (2 \pi)^{-D/2} \exp \left( - \frac{|S - \mu - b^* \theta^*|^2}{2} \right). \label{eq:eq_b}
\eea
The value of $\epsilon$ is found to be related to $a^*$, $b^*$, and $\theta^*$ via
\bea
\epsilon &=& \exp \left( a^* + b^* \mu^T \theta^* +  \frac{b^{*2} |\theta^*|^2}{2} \right). \label{eq:epsilon}
\eea
Appendix C provides an explicit derivation of \eq{eq:eq_b} and \eq{eq:epsilon}. 

We compute the distribution of model predictions for each bin as follows. For each bin $M$, this distribution is defined as
\bea
p(R|M) = \int dS\, \delta(R - \theta^T S) p(S|M).
\eea
This can be analytically calculated for both of the bins owing to the Gaussian form of each sequence distribution. We find that
\bea
p(R|M=0) &=& \frac{1}{\sqrt{2 \pi} |\theta|} \exp \left( - \frac{(R - \mu^T \theta)^2}{2 |\theta|^2} \right), \label{eq:prgm0} \\
p(R|M=1) &=& \frac{1}{\sqrt{2 \pi} |\theta|} \exp \left( - \frac{(R - [\mu + b^* \theta^*]^T \theta)^2}{2 |\theta|^2} \right).\label{eq:prgm1} 
\eea
See Appendix C for details. 

\subsection{Noise function}

To compute likelihood, we must posit a noise function $\pi(M|R)$. Based on our prior knowledge of the selection procedure, we choose $\pi(M|R)$ so that
\bea
\frac{\pi(M=1|R)}{\pi(M=0|R)} = \exp( a + bR),
\eea
where $a$ and $b$ are scalar parameters that we might or might not know \emph{a priori}. This, combined with the normalization requirement, $\sum_M \pi(M|R) = 1$, gives
\bea
\pi(M=1|R) = \frac{e^{a + bR}}{1 + e^{a + bR}},~~~~\pi(M=0|R) = \frac{1}{1 + e^{a + bR}}.\label{eq:gaussian_noise_function}
\eea
This noise function $\pi$ is correct when $a = a^*$ and $b = b^*$. The parameter $b$ is dual to the diffeomorphic mode $|\theta|$, whereas the parameter $a$ is not dual to any diffeomorphic mode.

In the experimental setup used to motivate the Gaussian selection model, the parameter $a$ is affected by many aspects of the experiment, including the concentration of the protein used in the binding assay, the efficiency of DNA extraction from the gel, and the relative amount of PCR amplification used for the bin 0 and bin 1 sequences. In practice, these aspects of the experiment are very hard to control, much less predict. From the results in the previous section, we can expect that if we assume a specific value for $a$ and perform likelihood-based inference, inaccuracies in this value for $a$ will distort our inferred model $\theta$ in an unpredictable (i.e., nondiffeomorphic) way. We will, in fact, see that this is the case. The solution to this problem, of course, is to  infer $\theta$ alone by maximizing the mutual information $I(\theta)$; in this case the values for $a$ and $b$ become irrelevant. Alternatively, one can place a prior on $a$ and $b$, then maximize noise-averaged likelihood $L_\mathrm{na}(\theta)$. We now analytically explore the consequences of these three approaches.

\subsection{Likelihood}

Using the noise function in \eq{eq:gaussian_noise_function}, the likelihood $L$ becomes a function of $\theta$, $a$, and $b$. Computing $L$ in the $N \to \infty$ and $\epsilon \to 0$ limits, we find that 
\bea
L(\theta,a,b) = \epsilon [ a + b \theta^T \mu + b b^* \theta^T \theta^* ] - \exp \left( a + b \theta^T \mu + \frac{b^2 |\theta|^2}{2} \right). \label{eq:gaussian_model_likelihood}
\eea

We now consider the consequences of various approaches for using $L(\theta,a,b)$ to estimate $\theta^*$. In each case, the inferred optimum will be denoted by a superscript `o.' Standard likelihood-based inference requires that we assume a specific value for $a$ and for $b$, then optimize $L(\theta, a, b)$ over $\theta$ alone by setting
\bea
0 &=& \left. \frac{\partial L}{\partial \theta_i} \right|_{\theta^o,a,b} 
\eea
for each component $i$. By this criteria we find that the optimal model $\theta^o$ is given by a linear combination of $\theta^*$ and $\mu$:
\bea
\theta^o = \frac{c b^*}{b} \theta^* + \frac{c-1}{b} \mu, \label{eq:likelihood_thetao}
\eea
where $c$ is a scalar that solves the transcendental equation 
\bea
c &=& \exp \left( [a^* - a] + \frac{1-c^2}{2}|b^* \theta^* + \mu|^2 \right).
\label{eq:sdf}
\eea
See Appendix C for the derivation of this result. Note that $c$ is determined only by the value of $a$ and not by the value of $b$. Moreover, $c = 1$ if and only if $a = a^*$. 

If our assumed noise function is correct, i.e., $a = a^*$ and $b = b^*$, then 
\bea
\theta^o = \theta^*.
\eea
Thus, maximizing likelihood will identify the correct model parameters. This exemplifies the general behavior illustrated in \fig{fig:hessians}{a}.

If $a = a^*$ but $b \neq b^*$, our assumed noise function will differ from the correct noise function only in a manner dual to the diffeomorphic mode $|\theta|$. In this case we find that $c = 1$ and
\bea
\theta^o = \frac{b^*}{b} \theta^*,
\eea
$\theta^o$ is thus proportional but not equal to $\theta^*$. This comports with our claim above that the diffeomorphic mode of the inferred model, i.e. $|\theta^o|$, will be biased so as to compensate for the error in the dual parameter $b$. This finding follows the behavior described in \fig{fig:hessians}{c}.

If $a \neq a^*$, however, $c \neq 1$.  As a result, $\theta^o$ is  a nontrivial linear combination of $\theta^*$ and $\mu$, and will thus point in a different direction than $\theta^*$. This is true regardless of the value of $b$. This behavior is illustrated in \fig{fig:hessians}{b}: errors in non-dual parameters of the noise function will typically lead to errors in nondiffeomorphic parameters of the activity model. 

We now consider the error bars that likelihood places on model parameters. Setting $\theta = \theta^o + \delta \theta$ and expanding $L(\theta,a,b)$ about $\theta^o$, we find that
\bea
N L(\theta, a^*, b^*) \approx N L(\theta^o, a^*, b^*) - \frac{N_1 b^{*2}}{2} \sum_{i,j} \Lambda_{ij} \delta \theta_i \delta \theta_j, \label{eq:hessian_expansion}
\eea
where $\Lambda_{ij} = \delta_{ij} + (\mu_i + b^* \theta^*_i)(\mu_j + b^* \theta^*_j).$ Note that all eigenvalues of $\Lambda$ are greater or equal to 1. Adopting the posterior distribution
\bea
p(\theta | \data) \sim e^{N L(\theta, a, b)}
\eea
therefore gives a covariance matrix on $\theta$ of
\bea
\braket{\delta \theta_i \delta \theta_j} = \frac{\Lambda^{-1}_{ij}}{N_1 b^{*2}}.
\eea
Thus, $\delta \theta \sim N_1^{-1/2}$ in all directions of $\theta$-space. Although $\Lambda_{ij}$ will change somewhat if $a$ and $b$ deviate from $a^*$ and $b^*$, this same scaling behavior will hold. Therefore, when the noise function is incorrect and $N$ is sufficiently large, the finite bias introduced into $\theta^o$ will cause $\theta^*$ to fall outside the inferred error bars.

\subsection{Mutual information}
In the $\epsilon \to 0$ limit, \eq{eq:mi} simplifies to
\bea
I(\theta) &=&  \epsilon \int dR\, p(R | M=1) \log  \frac{p(R|M=1)}{p(R|M=0)} + O(\epsilon^2). \label{eq:model_i_formula}
\eea
The lowest order term on the right hand side can be evaluated exactly using \eq{eq:prgm0} and \eq{eq:prgm1}:
\bea
I(\theta) &=& \frac{\epsilon b^{*2}}{2} \frac{(\theta^T \theta^*)^2}{|\theta|^2}. \label{eq:model_i_result}
\eea
See Appendix C for details.  Note that the expression on the right is invariant under a rescaling of $\theta$. This reflects the fact that $|\theta|$ is a diffeomorphic mode of the model defined in \eq{eq:gaussian_activity_model}.

To find the model $\theta^o$ that maximizes mutual information, we set
\bea
0 = \left. \frac{\partial I}{\partial \theta_i} \right|_{\theta^o} = \frac{\epsilon b^{*2} \theta^{oT} \theta^*}{|\theta^o|^2} \left[ \theta^*_i - \theta_i^o \frac{\theta^{oT} \theta^*}{|\theta^o|^2} \right]
\eea
The optimal model $\theta^o$ must therefore be parallel to $\theta^*$, i.e.
\bea
\theta^0 \propto \theta^*.
\eea
Expanding about $\theta = \theta^o + \delta \theta$ as above, we find that
\bea
N I(\theta) = N I(\theta^o) - \frac{N_1 b^{*2}}{2} (\delta \theta_\perp)^2 \label{eq:mi_constraints}
\eea
where $\delta \theta_\perp$ is the component of $\delta \theta$ perpendicular to $\theta^*$; see Appendix C. Therefore, if we use the posterior distribution $p(\theta | \data) \sim e^{N I(\theta)}$ to infer $\theta$, we find uncertainties in directions perpendicular to $\theta^*$ of magnitude $N_1^{-1/2}$. These error bars are only slightly larger than those obtained using likelihood, and have the same dependence on $N$. However, we find no constraint whatsoever on the component of $\delta \theta$ parallel to $\theta^*$. These results are illustrated by \fig{fig:hessians}{d}.

\subsection{Noise-averaged likelihood}

We can also compute the noise-averaged likelihood, $L_\mathrm{na}(\theta)$, in the case of a uniform prior on $a$ and $b$, i.e. $p(\pi) = p(a,b) = \mathcal{C}$ where $\mathcal{C}$  is an infinitesimal constant. We find that
\bea
\exp [ N L_\mathrm{na}(\theta) ] &=& \int d\pi\, p(\pi) \exp[N L(\theta,\pi)] \\
&=& \mathcal{C} \int_{-\infty}^\infty da\, \int_{-\infty}^\infty db\, \exp \left( N_1 [ a + b \theta^T \mu + b b^* \theta^T \theta^* ] - N \exp \left[ a + b \theta^T \mu + \frac{b^2 |\theta|^2}{2} \right]  \right) \label{eq:lna_integral_formula}\\
&=& \mathcal{C} \Gamma(N_1) \sqrt{\frac{2 \pi}{N |\theta|^2} } \exp \left( \frac{N_1 b^{*2}}{2} \frac{(\theta^T \theta^*)^2}{|\theta|^2} \right). \label{eq:lna_integral}
\eea
See the Appendix C for details. Thus, 
\bea
L_\mathrm{na}(\theta) = I(\theta) - \frac{1}{N} \log |\theta| + \const,
\eea
where the constant (which absorbs $\mathcal{C}$ entirely) does not depend on $\theta$. If we perform Bayesian inference using noise-averaged likelihood, i.e., using $p(\theta | \data) \sim e^{N L_\mathrm{na}(\theta)}$,
we will therefore find in the large $N$ limit that $\delta \theta_\perp$ is constrained in the same way as if we had used mutual information. The noise function prior we have assumed further results in weak constraints on $|\theta|$ that do not tighten as $N$ increases.\footnote{In the case at hand, $|\theta^o|$ is pushed all the way to zero. This is an artifact of the simple flat prior $p(a,b)$. If we instead adopt a weak Gaussian prior on $b$, we can still carry out the computation of $L_\mathrm{na}$ analytically, and in this case we find that $|\theta^o|$ is finite.} This is represented in \fig{fig:hessians}{e}. 

\section{Discussion}


The systematic study of quantitative sequence-function relationships in biology is just now becoming possible, thanks to the development of a variety of massively parallel experiments. Concepts and methods from statistical physics are likely to prove valuable for understanding this basic class of biological phenomena as well as for learning sequence-function relationships from data. 

In this paper we have discussed the problem of learning parametric models of sequence-function relationships from experiments having poorly characterized experimental noise. We have seen that standard likelihood-based inference, which requires an explicit model of experimental noise, will generally lead to incorrect model parameters due to errors in the assumed noise function. By contrast, mutual-information-based inference allows one to learn parametric models without having to assume any noise function at all. Mutual-information-based inference is unable to pin down the values of model parameters along diffeomorphic modes. This behavior reflects a fundamental difference between how diffeomorphic and nondiffeomorphic modes are constrained by data. Diffeomorphic modes arise from arbitrariness in the distinction between the activity model and the noise function. These findings were illustrated using an analytically tractable model for a massively parallel experiment.

The study of quantitative sequence-function relationships still presents many challenges, both theoretical and computational. One major practical difficulty with the mutual-information-based approach described here is the difficulty of accurately estimating mutual information from data. Although methods are available for doing this \cite{Khan:2007up}, it remains unclear whether any are accurate enough to enable computational sampling of the posterior distribution $p(\theta|\data) \sim e^{NI(\theta)}$, as suggested here. Moreover, none of these estimation methods is regarded as definitive. We believe this lack of clarity regarding how to estimate mutual information reflects the fact that the density estimation problem itself has never been fully solved, even in one or two dimensions. We are hopeful, however, that field-theoretic methods for estimating probability densities \cite{Bialek:1996wr,Kinney:2014el,Kinney:2015vu} might help resolve the problem of estimating low-dimensional probability densities as well as estimating mutual information. 

The problem of model selection poses a major theoretical challenge. Ideally, one would like to explore a hierarchy of possible model classes when fitting parametric models to data. However, when considering effective models it is unclear how to move far beyond independent site models (e.g., energy matrices) due to the number of parameters growing exponentially with the length of the sequence. Alternatively, when learning mechanistic models such as the model of the \emph{lac} promoter featured in Fig.\ 3, it is unclear how to go about systematically testing different arrangements of binding sites, different protein-protein interactions, and so on. We emphasize that this model prioritization problem is fundamentally theoretical, not computational, and as of now there is little clarity on how to address this matter. 

Finally, the geometric structure of sequence-function relationships presents an array of intriguing questions. For instance, very little is known (in any system) about how convex or glassy such landscapes in sequence space are, what their density of states looks like, etc.. Most of the biological and evolutionary implications of these aspects of sequence-function relationships also have yet to be worked out. We believe that the methods and ideas of statistical physics may lead to important insights into these questions in the near future. 

\section{Appendix A: Maximum likelihood under various noise functions}

At the correct noise function $\pi^*$, likelihood is given by
\bea
L(\theta,\pi^*) = \braket{\log \pi^*(M|R)}_\data
\eea
Taylor expanding this quantity about $\theta^*$ gives
\bea
L(\theta,\pi^*) = L(\theta^*,\pi^*) + \sum_i \left. \frac{\partial L}{\partial \theta_i} \right|_{\theta^*} (\theta_i - \theta^*_i) 
+ \frac{1}{2} \sum_{i,j} \left. \frac{\partial^2 L}{\partial \theta_i \partial \theta_j} \right|_{\theta^*} (\theta_i - \theta^*_i) (\theta_j - \theta^*_j) + \cdots .
\eea
We define the random vector $u$ in terms of the coefficient of the linear term of this expansion:
\bea
\frac{u_i}{\sqrt{N}} \equiv \left. \frac{\partial L}{\partial \theta_i} \right|_{\theta^*} &=& \left. \braket{\frac{\partial \log \pi^*(M|R)}{\partial R} \frac{\partial R}{\partial \theta_i}}_\data \right|_{\theta^*}.
\eea
Because $u_i/\sqrt{N}$ is defined as a sum of $N$ random terms, and because the mean of these terms vanishes, the covariance $\braket{u_i u_j}_\real$ will, by the central limit theorem, be given by
\bea
\braket{u_i u_j}_\real 
&=& \left. \braket{\left[ \frac{\partial \log \pi^*(M|R)}{\partial R} \right]^2 \frac{\partial R}{\partial \theta_i} \frac{\partial R}{\partial \theta_j}}_{S,M} \right|_{\theta^*}  \\
&=& \left. \sum_M  \int dR\, p(M,R) \left[ \frac{\partial \log \pi^*(M|R)}{\partial R} \right]^2 \braket{ \frac{\partial R}{\partial \theta_i} \frac{\partial R}{\partial \theta_j} }_{S|R,M} \right|_{\theta^*}. \label{eq:uu}
\eea

At $\theta = \theta^*$, Each measurement $M$ will provide no additional information about $S$ beyond that provided by the model prediction $R = \theta(S)$. Mathematically this means that
\bea
\left. p(S|R,M) \right|_{\theta^*} = \left. p(S|R) \right|_{\theta^*}
\eea
for all $S$, $R$, and $M$. Equivalently, the conditional expectation value of any sequence-dependent function $f(S)$ will obey
\bea
\left. \braket{f(S)}_{S|R,M} \right|_{\theta^*} = \left. \braket{f(S)}_{S|R} \right|_{\theta^*} \label{eq:same_expectations}
\eea
for all $M$. We use this fact to simplify \eq{eq:uu}:
\bea
\braket{u_i u_j}_\real &=& \left. \int dR\, p(R) \braket{ \frac{\partial R}{\partial \theta_i} \frac{\partial R}{\partial \theta_j} }_{S|R} \sum_M \pi(M|R) \left[ \frac{\partial \log \pi^*(M|R)}{\partial R} \right]^2 \right|_{\theta^*} \\
&=& \left. \int dR\, p(R) J(R) \braket{ \frac{\partial R}{\partial \theta_i} \frac{\partial R}{\partial \theta_j} }_{S|R}
\right|_{\theta^*} \label{eq:uu_2}
\eea
where $J(R)$ is the Fisher information from \eq{eq:fisher_information}.

We compute the Hessian of likelihood as follows:
\bea
H_{ij} 
\equiv \left. \frac{\partial^2 L(\theta,\pi^*)}{\partial \theta_i \partial \theta_j} \right|_{\theta^*} 
= \left. \braket{ \frac{\partial^2 \log \pi^*(M|R)}{\partial R^2} \frac{\partial R}{\partial \theta_i} \frac{\partial R}{\partial \theta_j}}_{S,M} \right|_{\theta^*}
+ \left. \braket{\frac{\partial \log \pi^*(M|R)}{\partial R} \frac{\partial^2 R}{\partial \theta_i \partial \theta_j} }_{S,M} \right|_{\theta^*}
\eea
The second term on the right hand side vanishes because of \eq{eq:same_expectations}:
\bea
\left. \braket{\frac{\partial \log \pi^*(M|R)}{\partial R} \frac{\partial^2 R}{\partial \theta_i \partial \theta_j} }_{S,M} \right|_{\theta^*} 
&=& \left. \sum_M \int dR\, p(R,M) \frac{\partial \log \pi^*(M|R)}{\partial R}\braket{\frac{\partial^2 R}{\partial \theta_i \partial \theta_j}}_{S|R,M} \right|_{\theta^*} \label{eq:start} \\
&=& \left. \int dR\, p(R) \braket{\frac{\partial^2 R}{\partial \theta_i \partial \theta_j}}_{S|R} \left[ \sum_M \pi(M|R) \frac{\partial \log \pi(M|R)}{\partial R} \right] \right|_{\theta^*} \\
&=& \left. \int dR\, p(R) \braket{\frac{\partial^2 R}{\partial \theta_i \partial \theta_j}}_{S|R} \left[ \frac{\partial}{\partial R} \sum_M \pi(M|R)  \right] \right|_{\theta^*} \\
&=& \left. \int dR\, p(R) \braket{\frac{\partial^2 R}{\partial \theta_i \partial \theta_j}}_{S|R} \left[ \frac{\partial}{\partial R} 1  \right] \right|_{\theta^*} \\
&=& 0. \label{eq:stop} 
\eea
We therefore find that
\bea
H_{ij} 
&=& \left. \sum_M \int dR\, p(R,M) \frac{\partial^2 \log \pi^*(M|R)}{R^2} \right|_{\theta^*}\\
&=& - \left. \int dR\, p(R) J(R) \braket{ \frac{\partial R}{\partial \theta_i} \frac{\partial R}{\partial \theta_i} }_{S|R} \right|_{\theta^*},
\eea
which is \eq{eq:likelihood_hessian_formula}. Note that, from \eq{eq:uu_2}, $\braket{u_i u_j}_\real = - H_{ij}$. 

The optimum $\theta^o$ of $L(\theta,\pi^*)$ will occur when
\bea
0 = \left. \frac{\partial L(\theta,\pi^*)}{\partial \theta} \right|_{\theta^o} = \frac{u_i}{\sqrt{N}} + \sum_j H_{ij} (\theta^o_i - \theta^*_i) + \cdots.
\eea
We therefore find that, to lowest order in $N^{-1/2}$,
\bea
\theta_i^0 = \theta_i^* - \sum_j H_{ij}^{-1} \frac{u_j}{\sqrt{N}}.
\eea
The covariance of $\theta^o$ is thus given by
\bea
\braket{(\theta_i^o - \theta_i^*)(\theta_j^o - \theta_j^*)}_\real = \sum_{k,l} H^{-1}_{ik} \frac{\braket{u_i u_j}_\real}{N} H^{-1}_{lj} = - \frac{H_{ij}^{-1}}{N},
\eea
which is \eq{eq:theta_o_cov}.

Under the incorrect noise function $\pi'$ defined in \eq{eq:noise_function_deviation},
\bea
L(\theta, \pi') &=& L(\theta, \pi^*) + \epsilon \braket{f(M,R)}_\data \\
&\approx& L(\theta^*,\pi^*) + \epsilon \left.  \braket{f(M,R)}_S\right|_{\theta^*} \nonumber \\ 
& & + \sum_i \left[ \frac{u_i}{\sqrt{N}}+ \epsilon w_i \right](\theta_i - \theta_i^*) 
 + \frac{1}{2} \sum_{ij} H_{ij} (\theta_i - \theta_i^*)(\theta_j - \theta_j^*)  + \cdots
\eea
where
\bea
w_i &=& \left. \braket{\frac{\partial f}{\partial R} \frac{\partial R}{\partial \theta_i} }_S \right|_{\theta^*}.
\eea
Let $\theta'$ denote the maximum of $L(\theta,\pi')$. Setting $\left. \frac{\partial L(\theta, \pi')}{\partial \theta_i} = 0 \right|_{\theta'}$, we find
\bea
\theta'_i = \theta_i^* - \sum_j H_{ij}^{-1} \left[ \frac{u_j}{\sqrt{N}} + \epsilon w_j \right], 
\eea
from which we get \eq{eq:deviation}.

In the case of a noise function $\pi''$ that differs from $\pi^*$ only along a dual mode, as in \eq{eq:dual_noise_function_deviation}, the vector $w_i$ is given by 
\bea
w_i 
&=& \left. \braket{\frac{\partial \tilde{v}^\mathrm{dif}}{\partial R} \frac{\partial R}{\partial \theta_i} }_S \right|_{\theta^*}.
\eea
The maximum likelihood parameters $\theta''$ will therefore satisfy
\bea
0 &=& \sum_j H_{ij}\braket{\theta''_j - \theta_j^*}_\real + \epsilon  w_i \\
&=& \sum_j \left.\braket{\frac{\partial^2 \log \pi}{\partial R^2} \frac{\partial R}{\partial \theta_i} \frac{\partial R}{\partial \theta_j} + \frac{\partial \log \pi}{\partial R} \frac{\partial^2 R}{\partial \theta_i \partial \theta_j}}_{S,M} \right|_{\theta = \theta^*} \braket{\theta''_j - \theta_j^*}_\real \nonumber \\
& & + 
\epsilon \left. \braket{  \frac{\partial R}{\partial \theta_i} \frac{\partial}{\partial R} \left[ \frac{\partial \log \pi}{\partial R} h(R) \right] }_{S,M} \right|_{\theta = \theta^*} 
\\
&=& \left. \braket{ \frac{\partial}{\partial \theta_i} \frac{\partial \log \pi}{\partial R} \left[   \sum_j \frac{\partial R}{\partial \theta_j} \braket{\theta''_j - \theta_j^*}_\real  + \epsilon h(R) \right]  }_{S,M} \right|_{\theta = \theta^*} \\
&=& \left. \braket{ \frac{\partial}{\partial \theta_i} \frac{\partial \log \pi}{\partial R}  \sum_j \frac{\partial R}{\partial \theta_j} \left(\braket{\theta''_j - \theta_j^*}_\real + \epsilon v_j^\mathrm{dif} \right)  }_{S,M} \right|_{\theta = \theta^*},
\eea
which is solved by \eq{eq:theta_prime_prime}. The fact that this uniquely specifies $\braket{\theta''_i - \theta^*_i}_\real$ follows from the Hessian $H$ being nonsingular.

\section{Appendix B: Gradient and Hessian of mutual information}

Here we calculate the gradient and Hessian of mutual information evaluated at $\theta = \theta^*$. We do this by first computing derivatives of the empirical probability distributions $p(R)$ and $p(R,M)$ with respect to model parameters. The mathematical trick used to do this is adapted from \cite{Sharpee:2004ty}. These results are first applied to likelihood in order to demonstrate their use and correctness. We then use this approach to compute the gradient and Hessian of mutual information. To clarify these derivations, we use $r(\theta,S)$, instead of $\theta(S)$, to explicitly denote the model prediction $R$ as a function of sequence $S$ and model parameters $\theta$. We also define $\partial_i \equiv \frac{\partial}{\partial \theta_i}$ and use $\int dS$ to represent sums over sequences. 

\subsection{How the distribution of model predictions changes with model parameters}

The empirical probability distribution of model predictions $R$ is given by
\bea
p(R) = \int dS\, p(S)\, \delta(R - r(\theta,S)).
\eea
The gradient of this probability distribution with respect to model parameters is computed as follows:
\bea
\partial_i p(R) 
&=&  \int dS\, p(S)\, \partial_i \delta(R - r(\theta,S)) \\
&=& - \int dS\, p(S) \left[ \frac{\partial}{\partial R} \delta(R - r(\theta,S) )\right] \partial_i r \\
&=& - \frac{\partial}{\partial R} \left[ p(R) \int dS\, p(S|R) \delta(R - r(\theta,S)) \partial_i r \right] \\
&=& - \frac{\partial}{\partial R} \left[ p(R) \braket{\partial_i r}_{S|R} \right].
\eea
Similarly, the Hessian of $p(R)$ is given by
\bea
\partial_i \partial_j p(R) 
&=& \int dS\, p(S) \left\{ 
\left[ \frac{\partial^2}{\partial R^2} \delta(R - r(\theta,S)) \right] \partial_i r \partial_j r 
- \left[ \frac{\partial}{\partial R} \delta(R - r(\theta,S)) \right] \partial_i \partial_j r 
\right\} \\
&=& \frac{\partial^2}{\partial R^2} \left[ p(R) \braket{\partial_i r \partial_j r}_{S|R} \right]
- \frac{\partial}{\partial R} \left[ p(R) \braket{\partial_i \partial_j r}_{S|R} \right].
\eea
Analogous results follow for the gradient and Hessian of the joint distribution $p(R,M)$:
\bea
\partial_i p(R,M) 
&=& - \frac{\partial}{\partial R} \left[ p(R,M) \braket{\partial_i r}_{S|R,M} \right] \\
\partial_i \partial_j p(R,M) 
&=& \frac{\partial^2}{\partial R^2} \left[ p(R,M) \braket{\partial_i r \partial_j r}_{S|R,M} \right]
- \frac{\partial}{\partial R} \left[ p(R,M) \braket{\partial_i \partial_j r}_{S|R,M} \right].
\eea

\subsection{Gradient and Hessian of likelihood}

Likelihood can be expressed in terms of the empirical distribution $p(R,M)$ as
\bea
L(\theta,\pi) &=& \sum_M \int dR\, p(R,M) \log \pi(M|R).
\eea
Keep in mind that $R$ is just a dummy variable in this integral; the empirical distribution $p$ is the only quantity that depends on $\theta$. The gradient of likelihood is therefore computed as
\bea
\partial_i L 
&=& \sum_M \int dR\, [ \partial_i p(R,M) ] \log \pi(M|R) \label{eq:likelihood_grad} \\
&=& \sum_M \int dR\, \left\{ - \frac{\partial}{\partial R} \left[ p(R,M) \braket{\partial_i r}_{S|R,M}\right]  \right\} \log \pi(M|R) \label{eq:L_grad_a} \\
&=& \sum_M \int dR\, p(R,M) \frac{\partial \log \pi(M|R)}{\partial R} \braket{\partial_i r}_{S|R,M} \label{eq:L_grad_b} \\
&=& \braket{ \frac{\partial \log \pi(M|R)}{\partial R} \partial_i r}_{S,M}. 
\eea
Note that in going from \eq{eq:L_grad_a} to \eq{eq:L_grad_b} we used integration by parts. The Hessian of likelihood is computed similarly:
\bea
\partial_i \partial_j L 
&=& \sum_M \int dR\, [ \partial_i \partial_j p(R,M) ] \log \pi(M|R) \label{eq:likelihood_hessian} \\
&=& \sum_M \int dR\, \log \pi(M|R) \left\{
\frac{\partial^2}{\partial R^2} \left[ p(R,M) \braket{\partial_i r \partial_j r}_{S|R,M} \right]
- \frac{\partial}{\partial R} \left[ p(R,M) \braket{\partial_i \partial_j r}_{S|R,M} \right]
\right\}  ~~~~~ \\
&=& \sum_M \int dR\, p(R,M) \left\{ 
\frac{\partial^2 \log \pi(M|R)}{\partial R^2} \braket{\partial_i r \partial_j r}_{S|R,M} 
+ \frac{\partial \log \pi(M|R)}{\partial R}  \braket{\partial_i \partial_j r}_{S|R,M} 
\right\} \label{eq:L_hessian_xxy}.
\eea
This expression is valid for all choices $\theta$ and $\pi$.

Restricting our attention now to $\theta = \theta^*$ and $\pi = \pi^*$, we see that the second term in \eq{eq:L_hessian_xxy} vanishes as it did in \eq{eq:start} through \eq{eq:stop}. Moreover, the first term gives
\bea
\partial_i \partial_j L = -\int dR\, p(R) J(R) \braket{\partial_i r \partial_j r}_{S|R},
\eea
which is the formula obtained for $H_{ij}$ in \eq{eq:likelihood_hessian_formula}. 

\subsection{Gradient and Hessian of mutual information}

The gradient and Hessian computations for mutual information are simplified by expressing mutual information in terms of its component entropies. We write 
\bea
I(\theta) &=& H_R(\theta) + H_M - H_{RM}(\theta)
\eea
where
\bea
H_{RM}(\theta) &=& -\sum_M \int dR\, p(R,M) \log p(M,R), \\
H_{R}(\theta) &=& -\int dR\, p(R) \log p(R), \\
H_{M} &=& -\sum_M p(M) \log p(M). 
\eea
The gradient of $H_R$ is given by
\bea
\partial_i H_R &=& -\int dR\, [\partial_i p(R)] \log p(R) - \int dR\, p(R) \partial_i \log p(R) \\
&=& -\int dR\, [\partial_i p(R)] \log p(R) - \int dR\, p(R) \frac{1}{p(R)} \partial_i p(R) \\
&=& -\int dR\, [\partial_i p(R)] \log p(R) -  \partial_i 1 \\
&=& -\int dR\, [\partial_i p(R)] \log p(R).
\eea
Similarly,
\bea
\partial_i H_{RM} &=& -\sum_M \int dR\, [\partial_i p(R,M)] \log p(R,M).
\eea
$H_M$ doesn't depend on $\theta$, so $\partial_i H_M = 0$. The resulting gradient of mutual information is
\bea
\partial_i I &=& \sum_M \int dR\, [\partial_i p(R,M)] \log p(R,M) -\int dR\, [\partial_i p(R)] \log p(R) \\
&=& \sum_M \int dR\, [\partial_i p(R,M)] \log \frac{p(R,M)}{p(R)} \\
&=& \sum_M \int dR\, [\partial_i p(R,M)] \log p(M|R).
\eea
Note from \eq{eq:likelihood_grad} that $\partial_i I = \partial_i L$ whenever $\pi(M|R) = p(M|R)$.

Now let's compute the Hessian of $H_R$:
\bea
\partial_i \partial_j H_R &=& -\int dR\, [\partial_i \partial_j p(R)] \log p(R) -\int dR\, [\partial_i p(R)] \partial_i \log p(R) \\
&=& -\int dR\, [\partial_i \partial_j p(R)] \log p(R) -\int dR\, p(R) [\partial_i \log p(R)] [\partial_j \log p(R)].
\eea
Similarly,
\bea
\partial_i \partial_j H_{RM} &=& - \sum_M \int dR\, [\partial_i \partial_j p(R,M)] \log p(R,M) \nonumber \\
& & - \sum_M \int dR\, p(R,M) [\partial_i \log p(R,M)] [\partial_j \log p(R,M)].
\eea

The Hessian of mutual information is therefore given by,
\bea
\partial_i \partial_j I = \partial_i \partial_j H_R - \partial_i \partial_j H_{RM}.
\eea
Using the form of $\partial_i \partial_j L$ in \eq{eq:likelihood_hessian}, we see that this reduces to
\bea
\partial_i \partial_j I = \partial_i \partial_j L + \Lambda_{ij}^{RM} - \Lambda_{ij}^R,
\eea
where
\bea
\Lambda^R_{ij} &=& \int dR\, p(R) [\partial_i \log p(R)] [\partial_j \log p(R)] 
\eea
and
\bea
\Lambda^{RM}_{ij} &=& \sum_M \int dR\, p(R,M) [\partial_i \log p(R,M)] [\partial_j \log p(R,M)].
\eea
We now split $\Lambda_{ij}^R$ and $\Lambda_{ij}^{RM}$ into four terms each. For $\Lambda_{ij}^R$ we get
\bea
\Lambda^R_{ij} &=& \int dR\, p(R) 
\left\{ - \frac{1}{p(R)}\frac{\partial}{\partial R} \left[ p(R) \braket{\partial_i r}_{S|R} \right] \right\} 
\left\{ - \frac{1}{p(R)}\frac{\partial}{\partial R} \left[ p(R) \braket{\partial_j r}_{S|R} \right] \right\} \\
&=& \int dR\, p(R) 
\left\{  \frac{\partial \log p(R)}{\partial R} \braket{\partial_i r}_{S|R} + \frac{\partial}{\partial R} \braket{\partial_i r}_{S|R} \right\} 
\left\{  \frac{\partial \log p(R)}{\partial R} \braket{\partial_j r}_{S|R} + \frac{\partial}{\partial R} \braket{\partial_j r}_{S|R} \right\} ~~\\
&=& A^R_{ij} + B^R_{ij} + B^R_{ji} + C^R_{ij},
\eea
where
\bea
A^R_{ij} &=& \int dR\, p(R) \left[ \frac{\partial \log p(R)}{\partial R} \right]^2 \braket{\partial_i r}_{S|R} \braket{\partial_j r}_{S|R}, \\
B^R_{ij} &=& \int dR\, p(R) \left[  \frac{\partial \log p(R)}{\partial R} \right] \braket{\partial_i r}_{S|R} \frac{\partial}{\partial R} \braket{\partial_j r}_{S|R}, \\
C^R_{ij} &=& \int dR\, p(R) \left[ \frac{\partial}{\partial R} \braket{\partial_i r}_{S|R} \right] \left[ \frac{\partial}{\partial R} \braket{\partial_j r}_{S|R} \right].
\eea
Similarly,
\bea
\Lambda^{RM}_{ij} &=& A^{RM}_{ij} + B^{RM}_{ij} + B^{RM}_{ji} + C^{RM}_{ij}
\eea
where
\bea
A^{RM}_{ij} &=& \sum_M \int dR\, p(R,M) \left[ \frac{\partial \log p(R,M)}{\partial R} \right]^2 \braket{\partial_i r}_{S|R,M} \braket{\partial_j r}_{S|R,M}, \\
B^{RM}_{ij} &=& \sum_M \int dR\, p(R,M) \left[  \frac{\partial \log p(R,M)}{\partial R} \right] \braket{\partial_i r}_{S|R,M} \frac{\partial}{\partial R} \braket{\partial_j r}_{S|R,M}, \\
C^{RM}_{ij} &=& \sum_M \int dR\, p(R,M) \left[ \frac{\partial}{\partial R} \braket{\partial_i r}_{S|R,M} \right] \left[ \frac{\partial}{\partial R} \braket{\partial_j r}_{S|R,M} \right].
\eea
It is unclear how to simplify the expression for $\partial_i \partial_j I$ at general choices of $\theta$. At $\theta = \theta^*$, however, the expectation value $\braket{\partial_i r}_{S|R,M}$ looses all $M$-dependence, and this causes a lot of cancellations to occur:
\bea
C^{RM}_{ij}
 &=& \sum_M \int dR\, p(R,M) \left[ \frac{\partial}{\partial R} \braket{\partial_i r}_{S|R} \right] \left[ \frac{\partial}{\partial R} \braket{\partial_j r}_{S|R} \right] \\
 &=& \int dR\, p(R) \left[ \frac{\partial}{\partial R} \braket{\partial_i r}_{S|R} \right] \left[ \frac{\partial}{\partial R} \braket{\partial_j r}_{S|R} \right] \\
 &=& C^R_{ij}
\eea
and
\bea
B^{RM}_{ij} 
&=& \int dR\, p(R) \left[\sum_M p(M|R)  \frac{\partial \log p(R,M)}{\partial R} \right]  \braket{\partial_i r}_{S|R}  \frac{\partial}{\partial R} \braket{\partial_j r}_{S|R} \\
&=& \int dR\, p(R) \left[\sum_M \frac{p(M|R)}{p(R,M)}  \frac{\partial p(R,M)}{\partial R} \right]  \braket{\partial_i r}_{S|R}  \frac{\partial}{\partial R} \braket{\partial_j r}_{S|R} \\
&=& \int dR\, p(R) \left[ \frac{1}{p(R)} \frac{\partial}{\partial R} \sum_M p(R,M) \right]  \braket{\partial_i r}_{S|R}  \frac{\partial}{\partial R} \braket{\partial_j r}_{S|R} \\
&=& \int dR\, p(R) \left[ \frac{1}{p(R)} \frac{\partial p(R)}{\partial R}  \right]  \braket{\partial_i r}_{S|R}  \frac{\partial}{\partial R} \braket{\partial_j r}_{S|R} \\
&=& \int dR\, p(R) \left[ \frac{\partial \log p(R)}{\partial R}  \right]  \braket{\partial_i r}_{S|R}  \frac{\partial}{\partial R} \braket{\partial_j r}_{S|R} \\
&=& B^R_{ij}.
\eea
We therefore find that,
\bea
\Lambda^{RM}_{ij} - \Lambda^R_{ij} &=& A_{RM} - A_R \\
&=& \int dR\, p(R) \braket{\partial_i r}_{S|R} \braket{\partial_j r}_{S|R} \left\{ \sum_M p(M|R) \left[ \frac{\partial \log p(R,M)}{\partial R} \right]^2  - \left[ \frac{\partial \log p(R)}{\partial R} \right]^2 \right\}~~~~
\eea
The expression in braces can be simplified as follows:
\bea
\sum_M p(M|R) \left[ \frac{\partial \log p(R,M)}{\partial R} \right]^2  &-& \left[ \frac{\partial \log p(R)}{\partial R} \right]^2  \nonumber \\
&=& \sum_M p(M|R) \left\{ \left[ \frac{\partial \log p(M|R)}{\partial R} +  \frac{\partial \log p(R)}{\partial R}\right]^2  - \left[ \frac{\partial \log p(R)}{\partial R} \right]^2 \right\} ~~~~\\
&=& \sum_M p(M|R) \left\{ \left[ \frac{\partial \log p(M|R)}{\partial R} \right]^2 +  \frac{\partial \log p(R)}{\partial R} \frac{\partial \log p(M|R)}{\partial R} \right\} \\
&=& J(R) + \frac{1}{p(R)} \frac{\partial p(R)}{\partial R} \frac{\partial }{\partial R} \sum_M p(M|R) \\
&=& J(R).
\eea
The Hessian of mutual information at $\theta = \theta^*$ therefore has a rather simple form:
\bea
K_{ij} = H_{ij} + \Lambda^{RM}_{ij} - \Lambda^R_{ij} = -\int dR\, p(R)\, J(R)\, \left[ \braket{\partial_i r \partial_j r}_{S|R} - \braket{\partial_i r}_{S|R} \braket{\partial_j r}_{S|R} \right],
\eea
which is \eq{eq:mi_hessian_formula}. 

\section{Appendix C: Gaussian selection model}

\subsection{Derivation of Eq.\ \ref{eq:epsilon} }

Applying Bayes's theorem twice,
\bea
p(S | M=1) = \frac{p(M=1|S)}{p(M=1)} p(S) = \frac{p(M=1|S)}{p(M=1)} \frac{p(M=0)}{p(M=0|S)} p(S | M=0).
\eea
Using Eqs.\ \ref{eq:null_library}, \ref{eq:selection_procedure}, and \ref{eq:epsilon_from_ps} then gives
\bea
p(S | M=1) &=& \epsilon^{-1} e^{a^* + b^* S^T \theta^*}  (2 \pi)^{-D/2} \exp \left( - \frac{|S - \mu|^2}{2} \right). \label{eq:eq_a}
\eea
Next we complete the square in the exponent:
\bea
- \frac{|S - \mu|^2}{2} + b^* S^T \theta^*  &=& - \frac{|S|^2 + |\mu|^2 - 2\mu^T S - 2 b^* S^T \theta^*}{2} \\
&=& - \frac{|S|^2 + |\mu|^2 + |b^* \theta^*|^2 - 2\mu^T S - 2 b^* S^T \theta^* + 2 b^* \mu^T \theta^*}{2} \nonumber \\
& & + \frac{|b^* \theta^*|^2}{2} + b^* \mu^T \theta^* \\
&=& - \frac{|S - \mu - b^* \theta|^2}{2} + \frac{|b^* \theta^*|^2}{2} + b^* \mu^T \theta^*. \label{eq:square_completed}
\eea
From the first term in \eq{eq:square_completed} we recover \eq{eq:eq_b}. To get $\epsilon$, we substitute \eq{eq:square_completed} into \eq{eq:eq_a}. Comparing this to \eq{eq:eq_b} then gives
\bea
1 = \epsilon^{-1} e^{a^*} \exp \left(\frac{|b^* \theta^*|^2}{2} + b^* \mu^T \theta^* \right)
\eea
Solving for $\epsilon$ recovers \eq{eq:epsilon}.

\subsection{Derivation of Eqs.\ \ref{eq:prgm0} and \ref{eq:prgm1} }
Here we describe how to  compute $p(R|M)$ where $R = \theta^T S$. We first consider the case of $M=0$. 
\bea
p(R | M=0)  &=& \int dS\, p(S|M=0) \delta ( R - S^T \theta) \\
&=& \int dS\, p(S|M=0) \delta ( [R - \mu^T \theta] - [S - \mu]^T \theta) \\
&=& \int dS\, p(S|M=0) \delta (R'- S'^T \theta) 
\eea
where $R' = R - \mu^T \theta$ and $S' = S - \mu$. We have chosen to work with $R'$ and $S'$ instead of $R$ and $S$ because $p(S'|M=0)$ is centered about 0. Now, split $S'$ up into the components parallel and perpendicular to $\theta$:
\bea
S' = S'_\perp + S'_\parallel \hat{\theta},
\eea
where $S'_\perp$ is a vector of dimension $D-1$ orthogonal to $\theta$, $S'_\parallel$ is a scalar, and $\hat{\theta} = \theta / |\theta|$. This definition gives ${S'}^\top \theta = S'_\parallel |\theta|$. Continuing with the integration,
\bea
p(R|M=0) &=& \int dS'_\perp \int_{-\infty}^{\infty} dS'_\parallel\, \delta(R' - S'_\parallel |\theta|) (2 \pi)^{-D/2} \exp \left( - \frac{S'^2_\perp}{2} - \frac{S'^2_\parallel}{2} \right) \\
&=& \int_{-\infty}^{\infty} dS'_\parallel\,  \delta(R' - S'_\parallel |\theta|) (2 \pi)^{-1/2} \exp \left( - \frac{S'^2_\parallel}{2} \right) \\
&=& \int_{-\infty}^{\infty} dS'_\parallel\,  \delta \left( \frac{R'}{|\theta|} - S'_\parallel \right) |\theta|^{-1} (2 \pi)^{-1/2} \exp \left( - \frac{S'^2_\parallel}{2} \right) \\
&=& |\theta|^{-1} (2 \pi)^{-1/2}  \exp \left( - \frac{R'^2}{2 |\theta|^2} \right).
\eea
Finally, substituting $R$ back for $R'$ gives
\bea
p(R|M=0) = \frac{1}{\sqrt{2 \pi} |\theta|} \exp \left( - \frac{(R - \mu^T \theta)^2}{2 |\theta|^2} \right).
\eea
To compute $p(R|M=1)$, we just replace $\mu \to \mu + b^* \theta^*$, giving
\bea
p(R|M=1) = \frac{1}{\sqrt{2 \pi} |\theta|} \exp \left( - \frac{(R - [\mu + b^* \theta^*]^T \theta)^2}{2 |\theta|^2} \right).
\eea

\subsection{Derivation of \eq{eq:gaussian_model_likelihood}}

We compute likelihood in the $N \to \infty$ limit as follows:
\bea
L(\theta, a, b) &=& \sum_M p(M) \int dR p(R|M) \log \pi(M|R) \\
&=& \frac{N_0}{N} \int dR\, p(R|M=0) \log \frac{1}{1 + e^{a+bR}} + \frac{N_1}{N} \int dR\, p(R|M=1) \log \frac{e^{a+bR}}{1 + e^{a+bR}}  \\
&\approx& - \frac{N_0}{N} \int dR\, p(R|M=0) e^{a+bR} + \frac{N_1}{N} \int dR\, p(R|M=1) [a + b R] \\
&\approx& - \braket{e^{a+bR}}_{S|M=0} + \epsilon \braket{a + b R}_{S|M=1}. \label{eq:almost_L}
\eea
In deriving \eq{eq:almost_L} we assumed that $e^{a + bR} \ll 1$ for all values of $R$ over which both $p(R|M=0)$ and $p(R|M=1)$ have significant support. This assumption necessarily holds in the $\epsilon \to 0$ limit. We have also kept only the lowest order terms in $\epsilon$. Note in particular that $\braket{e^{a+bR}}_{S|M=0}$ will be of order $\epsilon$.

The second term in \eq{eq:almost_L} can be directly read off from \eq{eq:prgm1}:
\bea
\braket{a + bR}_{S|M=0} = a + b\braket{R}_{S|M=0} = a + b \mu^T \theta + b b^* \theta^T \theta^*. \label{eq:second_term}
\eea
From \eq{eq:prgm0} we see that the first term in \eq{eq:almost_L} can be computed by completing the square:
\bea
-\frac{(R - \mu^T \theta)^2}{2|\theta|^2} + bR 
&=& - \frac{R^2 + (\mu^T \theta)^2 - 2 (\mu^T \theta) R - 2 b |\theta|^2 R}{2|\theta|^2} \\
&=& - \frac{R^2 + (\mu^T \theta)^2 + b^2 |\theta|^4 - 2 (\mu^T \theta) R - 2 b |\theta|^2  R + 2 (\mu^T \theta) b |\theta|^2}{2|\theta|^2} \nonumber \\
& & + b (\mu^T \theta) + \frac{b^2 |\theta|^2}{2} \\
&=& - \frac{(R - \mu^T \theta - b |\theta|^2)^2}{2|\theta|^2} + b (\mu^T \theta) + \frac{b^2 |\theta|^2}{2},
\eea 
from which we get
\bea
\braket{e^{a + bR}}_{S|M=1} = \exp \left[ a + b (\mu^T \theta) + \frac{b^2 |\theta|^2}{2} \right]. \label{eq:first_term}
\eea
Plugging \eq{eq:second_term} and \eq{eq:first_term} into \eq{eq:almost_L} gives the formula for $L(\theta,a,b)$ in \eq{eq:gaussian_model_likelihood}.

\subsection{Derivation of Eqs.\ \ref{eq:likelihood_thetao} and \ref{eq:sdf}}
Here we show how to derive the optimal $\theta$ for $L(\theta,a,b)$, with $a$ and $b$ fixed. Setting the gradient of $L$ with respect to $\theta$ to zero,
\bea
0 = \left. \frac{\partial L}{\partial \theta_i} \right|_{\theta^o}  &=& \epsilon b (\mu_i +  b^* \theta^*_i) - b(\mu_i + b \theta^o_i) \exp \left(a + b \mu^T \theta^o + \frac{b^2 |\theta^o|^2}{2} \right).
\eea
This gives
\bea
\mu_i + b \theta^o_i &=& \epsilon (\mu_i + b^* \theta_i^*) \exp \left(-a - b\mu^T \theta^o - \frac{b^2 |\theta^o|^2}{2} \right) \\
&=& c (\mu_i + b^* \theta_i^*)
\eea
where $c$ is a constant satisfying
\bea
c &=& \epsilon \exp \left(-a - b  \mu^T \theta^o - \frac{b^2 |\theta^o|^2}{2} \right) \\
&=& \exp \left( [a^* - a] + \mu^T[b^* \theta^* - b \theta^o] + \frac{b^{*2} |\theta^*|^2 - b^2 |\theta^o|^2}{2} \right). \label{eq:c_rhs}
\eea
We thus find \eq{eq:likelihood_thetao}. Note that the right hand side of the above equation depends implicitly on $c$ through the value of $\theta^o$. To eliminate $\theta^o$ from the equation for $c$, we let $\Lambda$ denote the $\theta^*$ dependent part of \eq{eq:c_rhs}, then substitute in \eq{eq:likelihood_thetao}: 
\bea
\Lambda &\equiv& \mu^T[b^* \theta^* - b \theta^o] + \frac{b^{*2}|\theta^*|^2 - b^2 |\theta^o|^2}{2} \\
&=& \mu^T[ b^* \theta^* (1-c) - (c-1) \mu] + \frac{b^{*2}|\theta^*|^2}{2} - \frac{|c b^* \theta^* + (c-1)\mu|^2}{2} \\
&=& (1-c)b^* \mu^T \theta^* + (1-c) |\mu|^2 + \frac{(1-c^2) b^{*2}|\theta^*|^2}{2} \nonumber \\
& & - \frac{(1-c)^2|\mu|^2}{2} - c(c-1)b^* \mu^T \theta^*
\eea
Using
\bea
(c-1) - c(c-1) = 1-c^2,~~~\mathrm{and}~~~(1-c)-\frac{(1-c)^2}{2} = \frac{1-c^2}{2},
\eea
we get
\bea
\Lambda &=& (1-c^2) b^* \mu^T \theta^* + \frac{(1-c^2) |\mu^2|}{2} + \frac{(1-c^2)b^{*2} |\theta^*|^2}{2}\\
&=& \frac{1-c^2}{2}|b^* \theta^* + \mu|^2.
\eea
We thus find the transcendental equation for $c$,
\bea
c &=& \exp \left( [a^* - a] + \frac{1-c^2}{2}|b^* \theta^* + \mu|^2 \right),
\eea
which is \eq{eq:sdf}.

\subsection{Derivation of \eq{eq:hessian_expansion}}

From the expression for likelihood in \eq{eq:gaussian_model_likelihood}, we find that the Hessian of likelihood is
\bea
H_{ij} &=& \left. \frac{\partial^2 L(\theta,a^*,b^*)}{\partial \theta_i \partial \theta_j} \right|_{\theta^*} \\
&=& [- b^{*2} \delta_{ij} - (b^* \mu_i + b^{*2} \theta_i^*)(b^* \mu_j + b^{*2} \theta_j^*)]\exp \left( a + b^* \theta^T \mu + \frac{b^{*2} |\theta|^2}{2} \right) \\
&=& - b^{*2} \epsilon \Lambda_{ij} \label{eq:asdfads}
\eea
where
\bea
\Lambda_{ij} \equiv \delta_{ij} + (\mu_i + b^* \theta_i^*)(\mu_j + b^* \theta_j^*).
\eea
Note: in deriving \eq{eq:asdfads} we used the expression for $\epsilon$ in \eq{eq:epsilon}. The expression in \eq{eq:hessian_expansion} further makes use of the approximations $N_1 \approx \epsilon N$, which will hold in the $\epsilon \to 0$ limit, and 
\bea
\left. \frac{\partial^2 L(\theta,a^*,b^*)}{\partial \theta_i \partial \theta_j} \right|_{\theta^o} \approx \left. \frac{\partial^2 L(\theta,a^*,b^*)}{\partial \theta_i \partial \theta_j} \right|_{\theta^*},
\eea
which will hold in the large $N$ limit.

\subsection{Derivation of Eqs.\ \ref{eq:model_i_formula} and \ref{eq:model_i_result}}
We derive \eq{eq:model_i_formula} as follows. To ease notation a bit, we define $p_M(R) = p(R | M)$. 
\bea
I[R;M] &=& \sum_{M = 0,1} \int dR\, p(M,R) \log \frac{p_M(R)}{p(R)} \\
&=& p(M=1) \int dR\, p_1(R) \log \frac{p_1(R)}{p(R)} + p(M=0) \int dR\, p_0(R) \log \frac{p_0(R)}{p(R)}   \\
&=& p(M=1) \int dR\, p_1(R) \log \frac{p_1(R)}{p_0(R)} + p(M=1) \int dR\, p_1(R) \log \frac{p_0(R)}{p(R)} \nonumber \\
& & + p(M=0) \int dR\, p_0(R) \log \frac{p_0(R)}{p(R)}  \\
&=&  p(M=1) \int dR\, p_1(R) \log \frac{p_1(R)}{p_0(R)} +  \int dR\, p(R) \log \frac{p_0(R)}{p(R)}. \label{eq:second_term_gets_ignored}
\eea
Because $p(M=1) = \epsilon + O(\epsilon^2)$, the first term in \eq{eq:second_term_gets_ignored} is the right hand side of \eq{eq:model_i_formula} to lowest order in $\epsilon$. We now show that the second term is of order $\epsilon^2$ and can therefore be ignored. Up to terms of order $\epsilon^2$, 
\bea
p(R) =(1 - \epsilon) p_0(R) +  \epsilon p_1(R).
\eea
Rearranging this gives
\bea
p_0(R) = \frac{p(R) - \epsilon p_1(R)}{1 - \epsilon}.
\eea
Plugging this into the second term of \eq{eq:second_term_gets_ignored} gives
\bea
\int dR\, p(R) \log \frac{p_0(R)}{p(R)} &=& \int dR\, p(R) \log \left[\frac{1}{1 - \epsilon} \left(1 - \epsilon \frac{p_1(R)}{p(R)} \right) \right] \\
&=& \int dR\, p(R) \log \left[1 + \epsilon \left( 1 - \frac{p_1(R)}{p(R)} \right) + O(\epsilon^2) \right] \\
&=& \epsilon \int dR\, p(R)  \left( 1 - \frac{p_1(R)}{p(R)} \right) + O(\epsilon^2) \\
&=& O(\epsilon^2).
\eea
 \eq{eq:model_i_result} is derived as follows:
\bea
I(\theta) &=& \epsilon \braket{\log \frac{p(R | M=1)}{p(R | M=0)}}_{M=1} \\
&=& \epsilon \braket{\frac{(R - \mu^T \theta)^2}{2 |\theta|^2} - \frac{([R - \mu^T \theta] - b^* \theta^T \theta^*)^2}{2 |\theta|^2}}_{M=1} \\
&=& \frac{\epsilon}{2 |\theta|^2} \braket{2 [R - \mu^T \theta] b^* \theta^T \theta^*  - (b^* \theta^T \theta^*)^2}_{M=1} \\
&=& \frac{\epsilon}{2 |\theta|^2} \left(2 [\braket{R}_{M=1} - \mu^T \theta] b^* \theta^T \theta^*  - (b^* \theta^T \theta^*)^2 \right) \\
&=& \frac{\epsilon}{2 |\theta|^2} \left(2 [b^* \theta^T \theta^*] b^* \theta^T \theta^*  - (b^* \theta^T \theta^*)^2 \right) \\
&=& \frac{\epsilon b^{*2}}{2} \frac{(\theta^T \theta^*)^2}{|\theta|^2}.
\eea

\subsection{Derivation of \eq{eq:mi_constraints}}

To derive \eq{eq:mi_constraints}, we set
\bea
\theta = \theta^* + \delta \theta_\parallel + \delta \theta_\perp
\eea
where $\delta \theta_\parallel$ is the deviation of $\theta$ in the direction of $\theta^*$, and $\delta \theta_\perp$ is the deviation perpendicular to $\theta^*$. This gives
\bea
\frac{(\theta^T \theta^*)^2}{|\theta|^2} &=& \frac{(|\theta^*|^2 + \delta \theta_\parallel^T \theta^*)^2}{|\theta^*|^2 + 2 \delta \theta_\parallel^T \theta^* + |\delta \theta_\parallel|^2 + |\delta \theta_\perp|^2} \\
&=& |\theta^*|^2 \frac{|\theta^*|^2 + 2 \delta \theta_\parallel^T \theta^* + |\delta \theta_\parallel|^2}{|\theta^*|^2 + 2 \delta \theta_\parallel^T \theta^* + |\delta \theta_\parallel|^2 + |\delta \theta_\perp|^2} \\
&=& |\theta^*|^2 \left( 1 - \frac{|\delta \theta_\perp|^2}{|\theta^*|^2} + \cdots \right) \\
&=& |\theta^*|^2 - |\delta \theta_\perp|^2 + \cdots.
\eea
The result in \eq{eq:mi_constraints} readily follows by substituting this into the formula for mutual information in \eq{eq:model_i_result}, then approximating the Hessian of mutual information at $\theta^o$ by the Hessian at $\theta^*$.

\subsection{Derivation of Eq.\ \ref{eq:lna_integral}}
Here we show how to evaluate the equation, \eq{eq:lna_integral_formula}, for the noise-averaged likelihood $e^{N L_\mathrm{na}(\theta)}$. First, interchange the order of integration and define $a' = a + b \theta^T \mu$. This gives,
\bea
e^{N L_\mathrm{na}(\theta)}  &=& \mathcal{C} \int_{-\infty}^{\infty} db \int_{-\infty}^{\infty} da'\, \exp \left[  N_1 \left[a' + b b^* \theta^T \theta^* \right] - N \exp \left( a' + \frac{b^2 |\theta|^2}{2} \right) \right].
\eea
Next, define $M = N \exp \left( \frac{b^2 |\theta|^2}{2} \right)$, $u = M e^{a'}$, and so $e^{a'} = u/M$, $e^{a'}da' = du/M$. This gives
\bea
e^{N L_\mathrm{na}(\theta)}  &=& \mathcal{C} \int_{-\infty}^{\infty} db\, e^{N_1 b b^* \theta^T \theta^*} \int_{-\infty}^{\infty} \left( e^{a'} da' \right) \left( e^{a'} \right)^{N_1-1}   \exp \left[ - M e^{a'} \right] \\
&=& \mathcal{C} \int_{-\infty}^{\infty} db\, e^{N_1 b b^* \theta^T \theta^*} M^{-N_1} \int_{0}^{\infty} du\, u^{N_1-1}   \exp[-u] \\
&=& \mathcal{C} \Gamma(N_1) \int_{-\infty}^{\infty} db\, e^{N_1 b b^* \theta^T \theta^*} M^{-N_1} \\
&=& \mathcal{C} \Gamma(N_1) \int_{-\infty}^{\infty} db\, \exp \left[ N_1 b b^* \theta^T \theta^* - N_1 \frac{b^2 |\theta|^2}{2} \right] \\
&=& \mathcal{C} \Gamma(N_1) \int_{-\infty}^{\infty} db\, \exp \left[ \frac{N_1 |\theta|^2}{2} \left(2  b b^* \frac{\theta^T \theta^*}{|\theta|^2} -  b^2  \right) \right] \\
&=& \mathcal{C} \Gamma(N_1) \int_{-\infty}^{\infty} db\, \exp \left[ \frac{N_1 |\theta|^2}{2} \left( \frac{b^{*2}(\theta^T \theta^*)^2}{|\theta|^4} - \left[ b - \frac{b^* \theta^T \theta^*}{|\theta|^2} \right]^2 \right) \right] \\
&=& \mathcal{C} \Gamma(N_1) \exp \left( \frac{N_1b^{*2}}{2} \frac{(\theta^T \theta^*)^2}{|\theta|^2} \right) \int_{-\infty}^{\infty} db\, \exp \left( \frac{N_1 |\theta|^2}{2} \left[ b - \frac{b^* \theta^T \theta^*}{|\theta|^2} \right]^2  \right) \\
&=& \mathcal{C} \Gamma(N_1) \sqrt{\frac{2 \pi}{N_1 |\theta|^2}} \exp \left( \frac{N_1b^{*2}}{2} \frac{(\theta^T \theta^*)^2}{|\theta|^2} \right),
\eea
which is \eq{eq:lna_integral}.

\begin{acknowledgements} 
We would like to thank L. Peliti, O. Revoire, and T. Mora for organizing this special issue. 
This work was supported by the Simons Center for Quantitative Biology at Cold Spring Harbor Laboratory and the Starr Cancer Consortium (I7-A723).
\end{acknowledgements}

\bibliographystyle{spphys}       
\bibliography{15_jsp}   

\end{document}